\documentclass[fleqn,usenatbib]{mnras}

\usepackage{newtxtext,newtxmath}
\usepackage{amsmath}
\usepackage{graphics,graphicx}
\usepackage{natbib}
\usepackage{setspace}
\usepackage{caption,setspace}
\newcommand{\angstrom}{\mbox{\normalfont\AA}}

\usepackage{mathtools}
\usepackage{enumitem}
\usepackage{lipsum}
\setlist[itemize]{leftmargin=*}

\usepackage{floatrow}
\usepackage[outercaption]{sidecap}    
\sidecaptionvpos{figure}{t}

\usepackage[T1]{fontenc}
\usepackage{ae,aecompl}
\pdfoutput=1

\title[The LOFAR view of the cosmic star formation history]{The LOFAR Two-metre Sky Survey: the radio view of the cosmic star formation history}

\author[R.K. Cochrane et al.]{R. K. Cochrane,$^{1,2}$\thanks{E-mail: rcochrane@flatironinstitute.org}
R. Kondapally,$^{3}$
P. N. Best,$^{3}$
J. Sabater,$^{4,3}$
K. J. Duncan,$^{3}$ 
D. J. B. Smith,$^{5}$ 
\newauthor
M. J. Hardcastle,$^{5}$ 
H. J. A. R\"{o}ttgering,$^{6}$
I. Prandoni,$^{9}$
P. Haskell,$^{5}$
G. G\"{u}rkan,$^{7,8}$
and G. K. Miley$^{6}$
\vspace{0.2cm}\\
$^{1}$Center for Computational Astrophysics, Flatiron Institute, 162 Fifth Avenue, New York, NY 10010, USA \\
$^{2}$Center for Astrophysics | Harvard \& Smithsonian, 60 Garden St. Cambridge, MA 02138, USA\\
$^{3}$Institute for Astronomy, University of Edinburgh, Royal Observatory, Blackford Hill, Edinburgh, EH9 3HJ, UK\\
$^{4}$UK Astronomy Technology Centre, Royal Observatory, Blackford Hill, Edinburgh, EH9 3HJ, UK\\
$^{5}$Centre for Astrophysics Research, University of Hertfordshire, College Lane, Hatfield, AL10 9AB, UK\\
$^{6}$Leiden Observatory, Leiden University, PO Box 9513, NL-2300 RA Leiden, The Netherlands\\
$^{7}$Th\"uringer Landessternwarte, Sternwarte 5, D-07778 Tautenburg, Germany\\
$^{8}$CSIRO Space and Astronomy, ATNF, PO Box 1130, Bentley WA 6102, Australia\\
$^{9}$INAF-IRA, Via P. Gobetti 101, I-40129 Bologna, Italy}
\date{Accepted 2023 May 22. Received 2023 May 22; in original form 2023 January 20}
\pubyear{2023}
\begin{document}
\label{firstpage}
\pagerange{\pageref{firstpage}--\pageref{lastpage}}
\maketitle

\begin{abstract}
We present a detailed study of the cosmic star formation history over $90$ per cent of cosmic time ($0\lesssim z\lesssim4$), using deep, radio continuum observations that probe star formation activity independent of dust. The Low Frequency Array Two Metre Sky Survey has imaged three well-studied extragalactic fields, Elais-N1, Bo{\"o}tes and the Lockman Hole, reaching $\sim20\,\mu\rm{Jy/beam}$ rms sensitivity at $150\,\rm{MHz}$. The availability of high-quality ancillary data from ultraviolet to far-infrared wavelengths has enabled accurate photometric redshifts and the robust separation of radio-bright AGN from their star-forming counterparts. We capitalise on this unique combination of deep, wide fields and robustly-selected star-forming galaxies to construct radio luminosity functions and derive the cosmic star formation rate density. We carefully constrain and correct for scatter in the $L_{150\,\rm{MHz}}-\rm{SFR}$ relation, which we find to be $\sim0.3\,\rm{dex}$. Our derived star formation rate density lies between previous measurements at all redshifts studied. We derive higher star formation rate densities between $z\sim0$ and $z\sim3$ than are typically inferred from short wavelength emission; at earlier times, this discrepancy is reduced. Our measurements are generally in good agreement with far-infrared and radio-based studies, with small offsets resulting from differing star formation rate calibrations.
\end{abstract}
\section{Introduction}
Characterising the history of cosmic star formation and understanding the drivers of star formation in galaxies from high to low redshift have been fundamental goals of extragalactic research for several decades \citep[see][for a thorough review]{Madau2014}. Early high-redshift surveys presented a broad view of star formation over cosmic time, with the volume-averaged star formation rate density (SFRD) increasing from high redshift to peak somewhere in the range $1 \lesssim z \lesssim 2.5$ and then declining towards the present day \citep{Lilly1995,Lilly1996,Madau1996,Connolly1997,Pascarelle1998}. By the mid-2000s, the SFRD had been constrained fairly tightly back to $z\sim1$, using a range of tracers \citep{Hopkins2006,Wilkins2008}. However, its form at higher redshift and the exact position of the peak remained less well determined.\\
\indent Since then, numerous studies have attempted to constrain the SFRD more tightly, particularly at higher redshifts, and ultraviolet (UV) studies have probed unobscured star formation back to $z\sim10$ \citep[e.g.][]{Bowler2015,Bowler2017,Bouwens2015a,Oesch2018,Bouwens2019} and beyond \citep[e.g.][]{Donnan2022}. This is motivated, in part, by the need for better constraints on the physics of reionization; at $z>5$, the SFRD determines the contribution of star-forming galaxies (SFGs) to the budget of ionizing photons \citep[see the review by][]{Stark2016}. However, galaxy selections based on unobscured emission (e.g. using the Lyman break or Lyman alpha emission line) are biased towards galaxies that are both young and fairly dust poor \citep{Shibuya2020}, and substantial corrections are required to scale the UV-derived SFRD and bring it into line with infrared (IR)-derived values where the two overlap (at $z<3$; \citealt{Madau2014}). Such corrections are subject to considerable uncertainties on the degree of dust obscuration in galaxies. This is particularly unconstrained in the early Universe \citep[see][and references therein]{Ma2019}, and possibly underestimated: recent ALMA observations of the dust continuum emission from Lyman Break Galaxies suggests that individual galaxies as distant as $z\sim8$ can harbour significant amounts of dust \citep{Watson2015,Laporte2017,Bowler2018}. There are methods of estimating dust corrections from UV observations alone, primarily via the empirical $\rm{IRX}-\beta$ relation between the ratio of the infrared luminosity to UV luminosity, $\rm{IRX}$, and the UV spectral slope, $\beta$ \citep{Meurer1999}. However, the considerable scatter in the relation, due to complex geometries, older stellar populations and different intrinsic extinction curves \citep[see][]{Popping2017a,Narayanan2018}, add uncertainty to such corrections. Cosmic variance is also a concern for rest-frame UV studies of the highest redshift galaxies \citep{Trenti2008,Driver2010,Ventou2017}, as the depth needed to find these faint objects often comes at the expense of area. \\
\indent An alternative tracer of star formation is rest-frame far-infrared (FIR) emission; since this is driven by the thermal output of dust heated by young stars, it is used to quantify star formation that is missed by the UV. This is particularly critical around the peak of cosmic star formation ($z\sim1-3$), where $\sim85$ per cent of the total star formation is obscured \citep{Dunlop2017}. While the large area surveys enabled by {\it{Herschel}} and the South Pole Telescope have discovered some extreme sources as distant as $z\sim7$ \citep[e.g.][]{Weiss2013,Strandet2017,Marrone2018,Casey2019}, deeper surveys that have the sensitivity to detect more typical sources tend to be limited to small areas of sky \citep[e.g.][]{Aravena2016b,Hatsukade2016,Dunlop2017,Franco2018}. Since the redshift distribution of SMGs observed at $\sim1\,\rm{mm}$ peaks around $z=2.0-2.5$ \citep{Chapman2005,Koprowski2014,Simpson2014,Danielson2017,Stach2019}, constraints on the abundance of dusty sources and their contribution to the SFRD at $z>4$ are few, due to the lack of detected sources \citep{Casey2018b}. Recent work has shown the promise of untargeted, longer wavelength surveys in isolating higher redshift sources \citep{Zavala2021,Casey2021a,Cooper2021,Manning2022}. \cite{Zavala2021} present a large (for mm), $184\,\rm{arcmin}^{2}$, $2\,\rm{mm}$ survey, from which they identify $13$ sources. Combining their new data with an empirically-based model, they show that dust-obscured star formation dominates the cosmic star formation rate budget to $z=4$, dropping to a $35$ per cent contribution at $z=5$, and $20-25$ per cent at $z=6-7$ (broadly in line with \citealt{Dunlop2017} and \citealt{Bouwens2020}).\\
\indent The small field of view imaged in a single ALMA pointing makes extending untargeted sub-millimeter surveys to degree scales technically challenging \citep{Chen2022a}. However, building samples of robustly-characterised star-forming galaxies is critical in order to answer key questions in galaxy evolution. These include understanding the timing and drivers of cessation of star-formation in different types of galaxies, which is fundamentally linked to a robust measurement of the cosmic star formation rate density as a function of redshift. We would also like to be able to constrain better the amount of unobscured versus obscured star-formation at different epochs, and how this varies across the galaxy population. This requires large statistical samples that span a broad range of cosmic time (i.e. from $z\sim0$ to well beyond the peak of cosmic star formation), as well as wide fields to overcome cosmic variance. \\
\indent The new generation of radio interferometers offer a unique opportunity to provide such samples and resolve these issues. Unlike at UV and optical wavelengths, light at radio wavelengths is unaffected by dust obscuration. Non-thermal emission from supernovae at centimetre wavelengths has been shown directly to be a delayed, indirect tracer of star formation \citep{Condon1992,Cram1998}. Relativistic electrons spiralling in weak magnetic fields emit synchotron radiation, characterised by a smooth spectrum ($f_{\nu}\propto \nu^{\alpha}$,where $\alpha\sim-0.7$) over a large wavelength range. The broad utility of synchotron emission as a probe of star formation is also supported by the tight far-infrared to radio correlation (FIRC), which has been shown by many to hold over several orders of magnitude in radio luminosity \citep[e.g.][]{VanderKruit1971,Ivison2010a,Sargent2010,Bourne2011,Delhaize2017,Read2018,Mccheyne2021}. The sensitive, dust-independent nature of radio continuum emission as a star-formation rate tracer has been capitalised on by previous work estimating the star-formation history \citep[e.g.][]{Haarsma2000,Seymour2008}, including the detailed studies made possible by VLA observations of the COSMOS field \citep{Schinnerer2007,Sargent2010,Schinnerer2010,Delhaize2017,Novak2017,Smolcic2017,Leslie2020,VanderVlugt2020}. One key advantage of the radio is its ability to probe star formation and AGN activity in both local and very distant galaxies (radio-loud sources have been discovered at redshifts as high as $z=6.82$; \citealt{McGreer2006,Saxena2018,Belladitta2020,Banados2021,Ighina2021a}). However, to probe the bulk of the star-forming population, we need highly sensitive observations, and the deepest radio surveys typically cover limited areas of sky ($2\,\rm{deg}^{2}$ in the case of VLA-COSMOS).\\
\indent Census studies are now becoming feasible with the current generation of radio telescopes, which can map large sky areas with high sensitivity and good angular resolution in an efficient manner. The International Low Frequency Array (LOFAR; \citealt{VanHaarlem2013}), is a large array of radio antennas, centred in the Netherlands but with antenna stations around Europe. The large primary beam (full-width
at half-maximum  $3.8\,\rm{deg}^{2}$ for stations in the Netherlands) enables $10\,\rm{deg}^{2}$ regions of sky to be mapped in a single pointing. Making use of this, the LOFAR Two Metre Sky Survey (LoTSS; \citealt{Shimwell2017,Shimwell2019,Shimwell2022}) project is adopting a multi-pronged approach to surveying the Northern sky at radio wavelengths. One strand of this is a wide-field survey of the whole Northern sky at $120-168\,\rm{MHz}$ and $\sim6''$ angular resolution (see \citealt{Shimwell2017,Shimwell2019,Shimwell2022,Duncan2019,Williams2019}). The second strand is a series of deep-field pointings known as the LoTSS Deep Fields.\\
\indent The LoTSS Deep Fields currently comprise deep observations of three well-studied Northern extragalactic fields: the European Large-Area ISO Survey-North 1 (Elais-N1; \citealt{Kessler1996,Oliver2000}), Bo{\"o}tes \citep{Jannuzi1999} and the Lockman Hole \citep{Lockman1986}, which are expected to reach eventual depths of $\sim10\mu\rm{Jy/beam}$ rms \citep{Best2021}. The first data release of these Deep Fields data reach $\sim20\,\mu\rm{Jy/beam}$ at $150\,\rm{MHz}$\footnote{The central frequency of the LoTSS Deep Fields data is $144\,\rm{MHz}$ in Bo{\"o}tes and Lockman Hole, and $146\,\rm{MHz}$ in Elais-N1, but for simplicity, we will refer to the frequency as $150\,\rm{MHz}$.}  \citep{Duncan2021,Kondapally2021,Tasse2021,Sabater2021}. The radio imaging has been accompanied by a detailed programme of source association and cross-identification \citep{Kondapally2021}, photometric redshift estimation \citep{Duncan2021} and host galaxy characterisation \citep{Best2021}. In this paper, we perform a detailed study of the radio view of cosmic star formation using the three LOFAR Deep Fields, Elais-N1, Bo{\"o}tes and the Lockman Hole. The sensitivity of our observations is comparable to that of the $3\,\rm{GHz}$ COSMOS-VLA survey (this reached $2.3\,\mu\rm{Jy/beam}$, which is equivalent to $\sim19\,\mu\rm{Jy/beam}$ at $150\,\rm{MHz}$, assuming a radio spectral index of $-0.7$). Our data cover a substantially larger area ($\sim26\,\rm{deg}^{2}$ of overlap with ancillary data, across the three fields), providing $>80,000$ radio-identified galaxies with optical counterparts. Together, the multi-wavelength catalogues we have constructed identify diverse populations of galaxies, out to $z\sim6$ \citep{Best2021}. A complementary paper, \cite{Kondapally2021a}, presents the cosmic history of low-excitation radio galaxies. \\
\indent The structure of this paper is as follows. In Section 
\ref{sec:data}, we first describe the multi-wavelength coverage of the three fields, and introduce the methods used to match radio sources with multi-wavelength counterparts. We briefly present an overview of the derivation of photometric redshifts and physical properties of the radio sources, as well as the separation of star-forming galaxies from AGN. We describe the methods used to construct luminosity functions and present the evolution of the $150\,\rm{MHz}$ luminosity functions of star-forming galaxies in Section \ref{sec:main_LF_sec}. In Section \ref{sec:L_sfr_calibration}, we construct star formation rate functions (SFRFs) from these luminosity functions, and compare these to SFRFs derived using SED-estimated star SFRs. We also estimate the scatter on the relation between $150\,\rm{MHz}$ luminosity and star formation rate. In Section \ref{sec:sfrd_results}, we construct the cosmic star formation history, from $z\sim0$ to $z\sim4$. We draw conclusions in Section \ref{sec:conclusions}.\\
\indent Throughout this paper, we use a $H_{0} = 70\,\rm{km}\,\rm{s}^{-1}\,\rm{Mpc}^{-1}$, $\Omega_{M} = 0.3$ and $\Omega_{\Lambda} = 0.7$ cosmology, along with a \cite{Chabrier2003} Initial Mass Function.

\section{The data: panchromatic observations of Elais-N1, Bo{\"o}tes and the Lockman Hole}\label{sec:data}
In this section, we present an overview of the LOFAR $150\,\rm{MHz}$ observations as well as the cross-matched UV-FIR photometric catalogues used in this work.

\subsection{Radio observations with LOFAR}
Elais-N1, Bo{\"o}tes and the Lockman Hole were observed by LOFAR at $150\,\rm{MHz}$ frequency with the HBA (high-band antenna) array, in a series of $8$-hour pointings. Observations of Elais-N1 (phase center 16h11m00s +55\textdegree00'00''; J2000) took place as part of cycles 0, 2 \& 4, with $164\,\rm{hr}$ of integration time in total. Observations of Bo{\"o}tes (phase center 14h32m00s +34\textdegree30'00'') were taken in cycles 3 \& 8, with $80\,\rm{hr}$ of integration time in total. The Lockman Hole (phase center 10h47m00s +58\textdegree05'00'') was observed in cycles 3 \& 10, with integration time summing to $112\,\rm{hr}$. All three fields were calibrated and imaged using Netherlands-only baselines, which gives rise to an angular resolution of $6''$ (note that imaging using international stations is also possible; see \citealt{Jackson2022,Morabito2022a,Sweijen2022}). The rms sensitivity reached at the pointing centre was $20\,\mu\rm{Jy/beam}$ for Elais-N1, $32\,\mu\rm{Jy/beam}$ for Bo{\"o}tes and $22\,\mu\rm{Jy/beam}$ for the Lockman Hole. Sensitivity decreases further from the pointing center; for each field, the area enclosed by the $30$ per cent power point is $\sim25\,\rm{deg}^{2}$.  A full description of these observations and the radio data reduction process is presented in \cite{Tasse2021} and \cite{Sabater2021}. \\
\indent Radio source extraction was performed on the Stokes I radio image using the Python Blob Detector and Source Finder (PyBDSF; \citealt{Mohan2015}). The final radio catalogue comprises $84,862$ sources in Elais-N1, $36,767$ sources in Bo{\"o}tes, and $50,112$ sources in the Lockman Hole \citep{Sabater2021,Tasse2021}. As discussed in the following section, we use a subset of these sources in the following analysis, limiting the sample to radio-identified sources that lie in regions of overlap with key optical and infrared surveys. 

\subsection{Multi-wavelength data}\label{sec:multiwav_data}
The three fields have different photometric coverage. Here, we review the ancillary catalogues generated and described fully by \cite{Kondapally2021}.
\subsubsection{Ultraviolet and optical data}
In all three fields, near-UV (NUV) and far-UV (FUV) data ($\lambda_{\rm{eff}}=1350$ \& $2800\angstrom$) are provided by the Galaxy Evolution Explorer (GALEX) space telescope (Deep Imaging Survey data release 6 \& 7; \citealt{Martin2005,Morrissey2007}).\\
\indent In Elais-N1 and the Lockman Hole, $u$-band data are drawn from the Spitzer Adaptation of the Red-sequence Cluster Survey (SpARCS; \citealt{Muzzin2009}), which used the Canada-France-Hawaii Telescope (CFHT). In Lockman Hole, the SpARCS data also adds images in the $u$, $g$, $r$, and $z$ bands, and there are additional observations from the Red Cluster Sequence Lensing Survey (RCSLenS; \citealt{Hildebrandt2016}) in the $g$, $r$, $i$, and $z$ bands. In Elais-N1, optical ($g$, $r$, $i$, $z$ \& $y$) broad-band imaging is provided by the Panoramic Survey Telescope and Rapid Response System (Pan-STARRS) 1 survey (PS1; \citealt{Chambers2016}). Further optical imaging from the Hyper-Suprime-Cam (HSC; \citealt{Aihara2018}) survey covers the United Kingdom Infrared Telescope (UKIRT) Infrared Deep Sky Survey (UKIDSS) Deep Extragalactic Survey (DXS) footprint (\citealt{Lawrence2007}; see below) with the broad-band filters $G$, $R$, $I$, $Z$ \& $Y$, as well as the narrow-band filter $\rm{NB}921$. In Bo{\"o}tes, deep optical photometry in the $B_{W}$, $R$ and $I$ bands is drawn from the NOAO Deep Wide Field Survey (NDWFS; \citealt{Jannuzi1999}), with additional $z$-band data from the zBo{\"o}tes survey \citep{Cool2007} and additional $U_{\rm{spec}}$ and $Y$-band imaging from the Large Binocular Telescope \citep{Bian2013}.

\subsubsection{Infrared data}
At near-infrared (NIR) wavelengths, the UKIDSS-DXS DR10, which used the UK Infrared Telescope (UKIRT), provides $J$ and $K$ band coverage  for Elais-N1 and the Lockman Hole. In Bo{\"o}tes, $J$, $H$ and $K_{S}$ data are drawn from the Infrared Bo{\"o}tes Imaging Survey, conducted with NEWFIRM on the Kitt Peak National Observatory Mayall 4-m telescope \citep{Gonzalez2010}.\\
\indent In the mid-infrared (MIR), {\it{Spitzer}}-IRAC observations at $3.6\,\mu\rm{m}$, $4.5\,\mu\rm{m}$, $5.8\,\mu\rm{m}$ and $8.0\,\mu\rm{m}$ are drawn from the {\it{Spitzer}} Wide-area Infra-Red Extragalactic (SWIRE; \citealt{Lonsdale2003}), which covers $\sim8\,\rm{deg}^{2}$ in Elais-N1 and $\sim11\,\rm{deg}^{2}$ in the Lockman Hole. In these two fields, we also draw data from the Spitzer Extragalactic Representative Volume Survey (SERVS) project \citep{Mauduit2012}, which covers $2.4\,\rm{deg}^{2}$ of Elais-N1 and $5.6\,\rm{deg}^{2}$ in the Lockman Hole, with the $3.6\,\mu\rm{m}$ and $4.5\,\mu\rm{m}$ channels, reaching $\sim1\,\rm{mag}$ deeper than SWIRE. In Bo{\"o}tes, data at $3.6\,\mu\rm{m}$, $4.5\,\mu\rm{m}$, $5.8\,\mu\rm{m}$ and $8.0\,\mu\rm{m}$ are primarily drawn from the {\it{Spitzer}} Deep, Wide-Field Survey (SDWFS; \citealt{Ashby2009}), with additional data from the Decadal IRAC Bo{\"o}tes Survey (M.L.N. Ashby PI, PID 10088).\\
\indent $24\,\mu\rm{m}$ data are provided by {\it{Spitzer}}-MIPS \citep{Rieke2004}; for Elais-N1 and the Lockman Hole, the data were drawn from the SWIRE survey. At longer FIR wavelengths, data for all three fields were drawn from the {\it{Herschel}} Multi-tiered Extragalactic Survey (HerMES; \citealt{Oliver2012}), which used the {\it{Herschel}} Space Observatory \citep{Pilbratt2010}. {\it{Herschel}} imaging at $100\,\mu\rm{m}$ and $160\,\mu\rm{m}$ comes from Photodetector Array Camera and Spectrometer (PACS; \citealt{Poglitsch2010}), and at $250\,\mu\rm{m}$, $350\,\mu\rm{m}$ and $500\,\mu\rm{m}$ from the Spectral and Photometric Imaging Receiver (SPIRE; \citealt{Griffin2010}). 
\subsubsection{Multi-wavelength catalogues}
The ultraviolet to mid-IR flux densities described above are compiled in the forced- and matched-aperture, aperture-corrected, multi-wavelength catalogues presented by \cite{Kondapally2021}. Sources were identified on combined chi-squared stack images; those with signal-to-noise ratio (SNR) less than 3 in all filters were removed from the catalogues.\\
\indent Full details of the FIR catalogues are provided in \cite{Mccheyne2021}, so we provide only a summary here. For all fields, FIR flux densities derived by the {\it{Herschel}} Legacy Project (HELP; \citealt{Shirley2021}) provided the basis of our measurements. As part of HELP, source deblending was performed using the Bayesian tool XID+ \citep{Hurley2017}. Where a LOFAR source could be cross-matched with a HELP catalogue entry within $0.5''$, the HELP fluxes were adopted. Where a LOFAR source had no HELP match, XID+ was re-run with the radio host galaxy position added to the prior list.

\subsection{A cross-matched radio and photometric catalogue}
Full details of the cross-matching of radio-identified sources to the photometric catalogues summarised in Section \ref{sec:multiwav_data} are provided in \cite{Kondapally2021}. At the depths reached by the LOFAR imaging in the Deep Fields, radio emission from multiple sources can be incorrectly linked by the PyBDSF source extraction procedure, leading to `blended sources'. The opposite scenario, extended emission from individual sources being split up into many components, can also occur. In this section, we summarise the approach employed to associate radio and multi-wavelength sources within the LOFAR Deep Fields. This was only performed over the area of sky with the best multi-wavelength data: $\sim6.74\,\rm{deg}^{2}$ in Elais-N1, $\sim8.63\,\rm{deg}^{2}$ in Bo{\"o}tes, and $\sim10.28\,\rm{deg}^{2}$ in the Lockman Hole.\\
\indent The likelihood ratio (LR) method \citep{DeRuiter1977,Sutherland1992} uses magnitude and colour information to match radio sources with their optical and IR counterparts, as described in \cite{Kondapally2021}. This technique yields robust associations for the majority of radio sources ($80-85$ per cent of sources in each field). For sources with extended or complex radio emission, sources were visually inspected using LOFAR Galaxy Zoo (LGZ; see also \citealt{Williams2019} for an earlier use of the same framework), a private Zooniverse project accessible to members of the LOFAR consortium. Other sources, primarily those that were potential radio blends, were sent for identification by an `expert user', who de-blended the PyBDSF source using the PyBDSF Gaussian components. \\
\indent Within the overlap region of PanSTARRS, UKIDSS and SWIRE in Elais-N1, there are $1,470,968$ optically-detected sources and $31,610$ radio-detected sources. After cross-matching to optical/NIR counterparts and the flagging of spurious sources, there were $30,839$ sources. Within the overlap region of the NDWFS and the SDWFS in Bo{\"o}tes, there are $1,911,929$ optically-detected sources, $19,179$ radio-detected sources, and $18,579$ sources in the final cross-matched catalogue. Within the overlap region of the SpARCS r-band and SWIRE in the Lockman Hole, there are $1,906,317$ optically-detected sources, $31,162$ radio-detected sources, and $30,402$ sources in the cross-matched catalogue. As discussed in some detail by \cite{Kondapally2021}, $2-3$ per cent of sources ($771$ in Elais-N1, $600$ in Bo{\"o}tes, and $760$ in the Lockman Hole) have no robust optical/NIR ID. Some of these sources have FIR counterparts, and are likely dusty AGN and SFGs. Note that \cite{Novak2017} reported that $\sim4$ per cent of VLA-COSMOS sources are optically faint shortward of $i$-band for a similar radio-selected population. NIR/optically-dark sources have also been reported as common in sub-millimeter selected samples \citep[e.g.][]{Simpson2014,Simpson2016,Franco2018,Wang2019a,Dudzeviciute2019,Smail2020}. \cite{Kondapally2021} investigated the LOFAR-detected radio sources without IDs, and found that these are likely dominated by $z\sim2-4$ AGN. Since we focus on star forming galaxies in this paper, they are unlikely to have much effect on our analysis.

\begin{figure*}
\includegraphics[width=6.82cm]{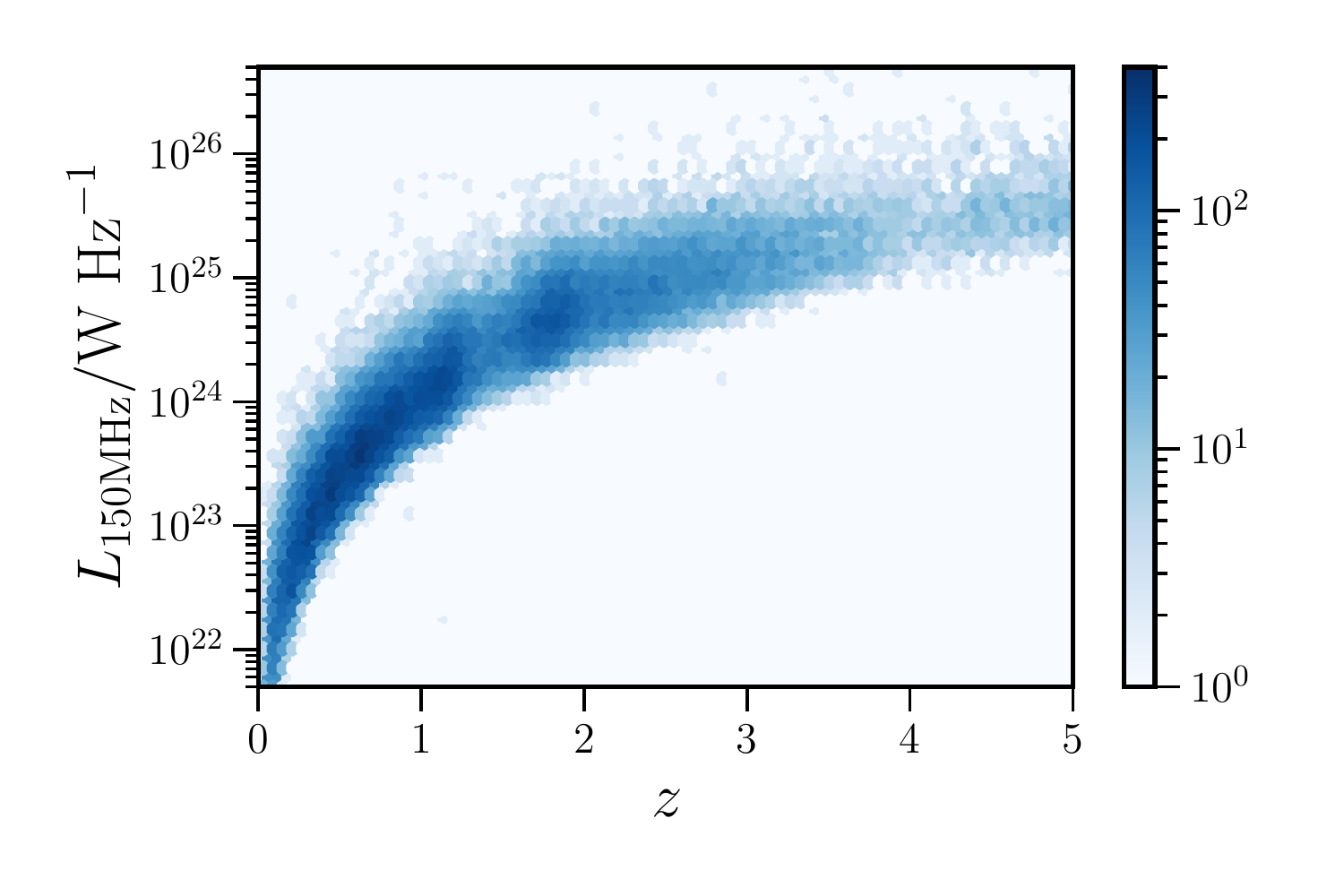}
\hspace{-1.45cm}
\includegraphics[width=6.82cm]{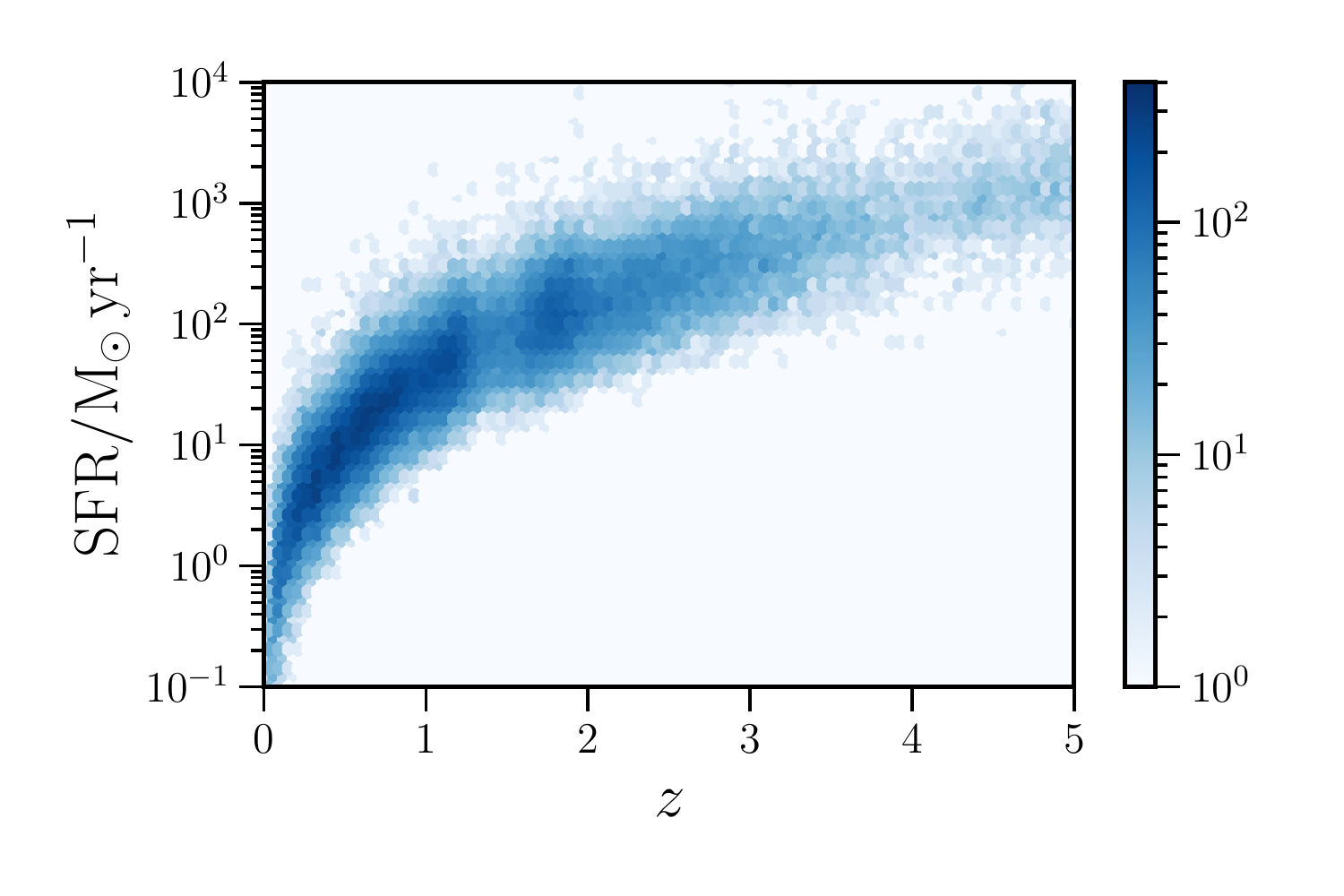}
\hspace{-1.45cm}
\includegraphics[width=6.82cm]{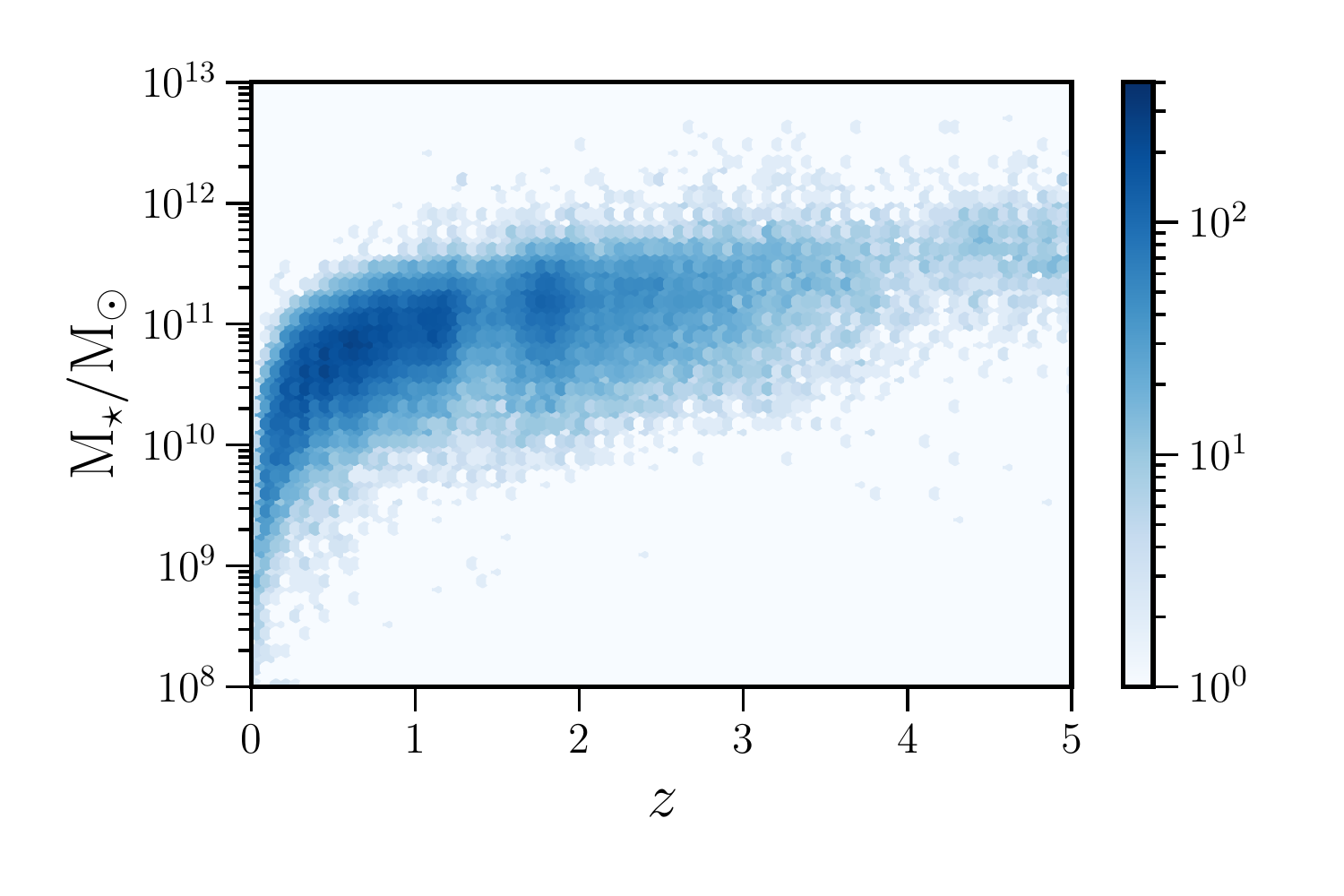}
\caption[]{The distribution of $150\,\rm{MHz}$ radio luminosity, star formation rate and stellar mass, for the main sample of galaxies studied in this paper (all galaxies within the three fields for which radio continuum emission at $150\,\rm{MHz}$ is dominated by star formation; see Section \ref{sec:sf_vs_agn} for details of the SFG/AGN separation and sample selection). The redshifts plotted are $z_{\rm{BEST}}$ - the spectroscopic redshift where this is available, and the median redshift of the preferred photometric redshift solution, $z_{\rm{1,MEDIAN}}$, where it is not. The stellar mass and SFR values plotted are the consensus estimates derived from the combination of four different SED fitting codes (see Section \ref{sec:sfrs_masses} and \citealt{Best2021}).}
\label{fig:demographics}
\end{figure*}

\subsection{Deriving photometric redshifts of LOFAR-identified sources}\label{sec:photzs}
The process of deriving photometric redshifts for sources in all three fields is described fully in \cite{Duncan2021} and we review only the most important details here. Both template fitting and machine learning techniques were used to derive photometric redshifts for both the radio-selected LOFAR sources, and the full optical catalogue ($\sim5$ million sources in total). This `hybrid' approach was shown by \cite{Duncan2018} to improve upon traditional template fitting, particularly for intermediate redshift AGN, which had proved a challenging population to obtain redshifts for \citep{Duncan2017}. \cite{Duncan2021} showed that outlier fractions, defined as $|z_{\rm{phot}}-z_{\rm{spec}}|/(1+z_{\rm{spec}})>0.15$, range from $1.5$ to $1.8$ per cent for galaxies, and from $18$ to $22$ per cent for optical, IR and X-ray selected AGN. In this paper, we make use of the photometric redshift posteriors for sources without spectroscopic redshifts (this is the majority; only $\sim8.6$ per cent of all LoTSS Deep Fields radio sources have spectroscopic redshifts). Since we exclude the majority of AGN from our analysis, these redshifts are reasonably robust ($\lesssim 2$ per cent outlier fractions).

\subsection{Classification of star-forming galaxies and AGN}\label{sec:sf_vs_agn}
Reliable classifications of star-forming sources and AGN are essential for this study. \cite{Best2021} present the multiple methods used to identify AGN from emission in different wavebands, which we summarise here. A combination of spectral energy distribution (SED) fitting codes was used to identify the majority of radiative-mode AGN (i.e. those identifiable via their optical/IR/X-ray emission). Each radio-identified source was fitted with four different SED fitting codes: {\small{MAGPHYS}} \citep{DaCunha2008}, {\small{BAGPIPES}} \citep{Carnall2018,Carnall2019}, {\small{AGNfitter}} \citep{Rivera2016}, and {\small{CIGALE}} \citep{Burgarella2005,Noll2009,Boquien2019}; see Section \ref{sec:sfrs_masses} for more details. The use of several different codes, two that include AGN templates ({\small{AGNfitter}} and {\small{CIGALE}}) and two that are optimised for `normal' galaxies, with more flexibility to fit star-forming populations ({\small{MAGPHYS}} and {\small{BAGPIPES}}), enables optimised fitting for the different radio populations. {\small{AGNfitter}} and {\small{CIGALE}} both constrain an $f_{\rm{AGN}}$ parameter, which describes the fractional AGN contribution to the IR emission (over the wavelength range $\sim5-1000\mu\rm{m}$ for {\small{CIGALE}} and the range $\sim1-30\mu\rm{m}$ for {\small{AGNfitter}}). Sources that were assigned a high value of $f_{\rm{AGN}}$ by both {\small{AGNfitter}} and {\small{CIGALE}} were classified as radiative-mode AGN, as were those that were assigned a high $f_{\rm{AGN}}$ by one of these codes and were also substantially better fitted (as assessed by the reduced $\chi^{2}$ values) by one of these codes than by {\small{MAGPHYS}} or {\small{BAGPIPES}} (suggesting that an AGN component was necessary in the fit).\\
\indent Additional optical AGN were identified via cross-matching to the Million Quasar catalogue, which is mainly based on the Sloan Digital Sky Survey \citep{Alam2015}. X-ray AGN (defined by X-ray-to-optical flux ratios or hardness ratios) were identified in Bo{\"o}tes thanks to the deep X-ray data provided by Chandra, as part of the X-Bo{\"o}tes survey of NDWFS \citep{Kenter2005}. In Elais-N1 and the Lockman Hole, the Second {\it{ROSAT}} All-Sky Survey \citep{Boller2016} and the {\it{XMM-Newton}} Survey provide shallower data, enabling the identification of brighter X-ray sources.\\
\indent Radio-selected AGN were identified based on their excess emission at $150\,\rm{MHz}$. For each source, the radio emission expected for a purely star-forming source with a given SED-derived SFR (see Section \ref{sec:sfrs_masses}) was determined using the `ridgeline' relation between SFR and $L_{150\,\rm{MHz}}$ derived by \cite{Best2021}; note that the `ridgeline' agrees well with the relation derived by \cite{Smith2021}. Then, excess radio emission relative to this expected value was calculated. Sources that were either offset from the relation by more than $0.7\,\rm{dex}$, or had extended ($>80\,\rm{kpc}$) radio emission, are likely not powered solely by star formation, and were hence classified as radio-selected AGN.\\
\indent In our final classifications, star-forming galaxies were defined as those without AGN signatures in optical/IR/X-ray, and without excess radio emission. Sources with excess radio emission form the radio-loud AGN classes. In this paper, we focus on galaxies without radio excess, i.e. galaxies classified as either star-forming or radio-quiet AGN. These comprise $77\%$ of the total LoTSS sample ($81\%$ of sources with optical/NIR counterparts and classifications; see \citealt{Best2021}). For the radio-quiet AGN, star-formation is expected to drive the majority of the radio continuum emission detected by LOFAR (though there will be a minority of cases where a weak jet dominates; see \citealt{Macfarlane2021} for a model of quasar radio luminosity that comprises contribution from star formation and an AGN jet. They concluded that the jet-launching mechanism operates in all quasars, but with different efficiency. See also \citealt{Gurkan2019}).\\
\indent Our final sample comprises $55,581$ radio sources within the redshift range $0.1<z<5.7$ for which radio emission is dominated by star formation. $21,638$ of these are in Elais-N1, $12,787$ are in Bo{\"o}tes, and $21,156$ are in the Lockman Hole.

\subsection{Stellar masses and star formation rates for star-forming galaxies}\label{sec:sfrs_masses}
As noted in Section \ref{sec:sf_vs_agn}, four different SED fitting codes are used to fit all radio-selected sources. SED fitting provides an alternative way to estimate galaxy SFRs, compared with single wavelength flux calibrations. `Energy balance' based SED fitting ensures that the unobscured and obscured components of galaxy emission are fitted simultaneously and self-consistently with a combination of simple stellar populations, a dust model, and models of star formation history and galactic chemical evolution. Physical parameters may then be derived from the best-fitting combination of models. These codes have been tested on simulated galaxies \citep[e.g.][]{Hayward2014b,Dudzeviciute2019}. Each of the four SED fitting codes provides estimates for the physical properties of the sources, including stellar mass and star-formation rates. \cite{Best2021} describe the process by which these SED fitting outputs are combined to generate `consensus' estimates. \\
\indent For all sources that were not classified as radiative-mode AGN, stellar mass and SFR were both generally calculated using the mean of the logarithm of the values derived using {\small{MAGPHYS}} and {\small{BAGPIPES}}, providing both codes yielded a statistically acceptable fit (note that this was the case for $\sim90$ per cent of these sources). Where one fit was unacceptable, the estimate from the acceptable fit was adopted. For the radio-quiet AGN in our sample, stellar masses and SFRs were derived using more appropriate SED fitting techniques, which included a component of emission from the AGN. For the vast majority of these ($>94$ per cent), stellar masses and SFRs were derived using {\small{CIGALE}} (see \citealt{Best2021} for full details of the small number of cases where {\small{AGNfitter}} was preferred).\\
\indent The distributions of $150\,\rm{MHz}$ luminosity, star formation rate and stellar mass, for our selected population of star-forming galaxies and radio-quiet AGN, are shown in Figure \ref{fig:demographics}.
\section{The radio luminosity function of star-forming galaxies}\label{sec:main_LF_sec}
\subsection{Constructing the $\mathbf{150}\,\rm{\bf{MHz}}$ luminosity function}\label{sec:details_lf}
With our sample of star-forming galaxies and radio-quiet AGN in-hand, we construct $150\,\rm{MHz}$ luminosity functions at a range of redshifts. Approximately $90$ per cent of the radio sources in our sample are classified as `pure' SFGs, but we include radio-quiet AGN to perform a complete census of star formation (see also \citealt{Bonato2021}, who construct radio luminosity functions for the separate populations). Rest-frame $150\,\rm{MHz}$ luminosities are calculated using the 
following formula, where $\nu$ is the observed-frame frequency, $150\,\rm{MHz}$, $\alpha$ is the radio spectral index (we assume $\alpha=-0.7$), and $S_{\nu}$ is the flux density at the observed frequency. $z$ is the source redshift (we use the spectroscopic redshift, if it exists, and if not, the photometric redshift, as described in Section \ref{sec:photzs}), and $D_{L}$ is the corresponding luminosity distance:

\begin{equation}
    L_{\nu}=\frac{4\pi D_{L}^{2}(z)}{(1+z)^{1+\alpha}}S_{\nu}.
\end{equation}

\indent We aim to calculate the space density of sources per luminosity bin, the `luminosity function', as a function of redshift. First, we divide the sources into eleven wide redshift bins. We place the observed LOFAR sources into these redshift bins, and calculate a rest-frame $150\,\rm{MHz}$ luminosity for each source. We bin sources further, by rest-frame $150\,\rm{MHz}$ luminosity, and then for each luminosity bin at each redshift, we derive the $150\,\rm{MHz}$ luminosity function, $\Phi(L,z)$, using the non-parametric $1/V_{\rm{max}}$ method \citep{Schmidt1968}:
\begin{equation}\label{eq:LF}
\Phi(L,z) = \frac{1}{\Delta \log_{10} L} \sum_{i=1}^{N} \frac{1}{V_{{\rm{max}},i}},
\end{equation}
where $\Delta \log_{10} L$ is the width of the luminosity bin and $V_{\rm{max}}$ is the volume within the redshift bin over which a source with a given rest-frame $150\,\rm{MHz}$ luminosity would be observable, given the sensitivity limits of the survey. This is particularly important for a radio survey such as ours, where the depth is not uniform across each individual field, nor between fields. $V_{\rm{max}}$ is calculated for each source using:
\begin{equation}\label{eq:lf_vmax}
V_{{\rm{max}},i} =  \int_{z_{\rm{min}}}^{z_{\rm{max}}} V(z)\,\theta(S,z) \,dz ,
\end{equation}
where $V(z)\,dz$ is the whole-sky co-moving volume in the redshift range $[z, z+dz]$ and $\theta(S,z)$ is the fractional area  over which a source of that flux density would have been detected with $5\sigma$ signal-to-noise. We perform the integral numerically, by dividing the wide redshift bin (with edges $z_{\rm{min}}$ and $z_{\rm{max}}$) into narrow redshift slices of size $\Delta(z)=0.0001$. For each narrow redshift slice, we calculate the $150\,\rm{MHz}$ flux density that we would observe if the source were located at the centre of that redshift slice (using its known rest-frame luminosity). We then calculate $\theta(S,z)$ for that flux density using:
\begin{equation}\label{eq:theta_s_z}
\theta(S,z) = \frac{\Omega[S(z)]}{4\pi}\times C_{\rm{radio}}[S(z)]\times C_{\rm{photometric}}(z),
\end{equation}
where $\Omega[S(z)]$ is the solid angle over which a source with flux density $S$ can be detected at $5\sigma$, and $C_{\rm{radio}}[S(z)]$ is the radio completeness as a function of flux density, as derived in Section \ref{sec:radio_completeness}. $C_{\rm{photometric}}(z)$ is an order-unity correction factor which accounts for inaccuracies (such as aliasing) in the photometric redshifts, as derived in Section \ref{sec:photo_z_uncertainty}. This enables us to fold in the spatially-varying radio depths over the fields, as well as uncertainties in photometric redshifts. \\
\indent Repeating the process for each of the wide redshift bins enables us to construct luminosity functions from $z\sim0$ to $z\sim5$. 

\begin{figure}
\includegraphics[width=\columnwidth]{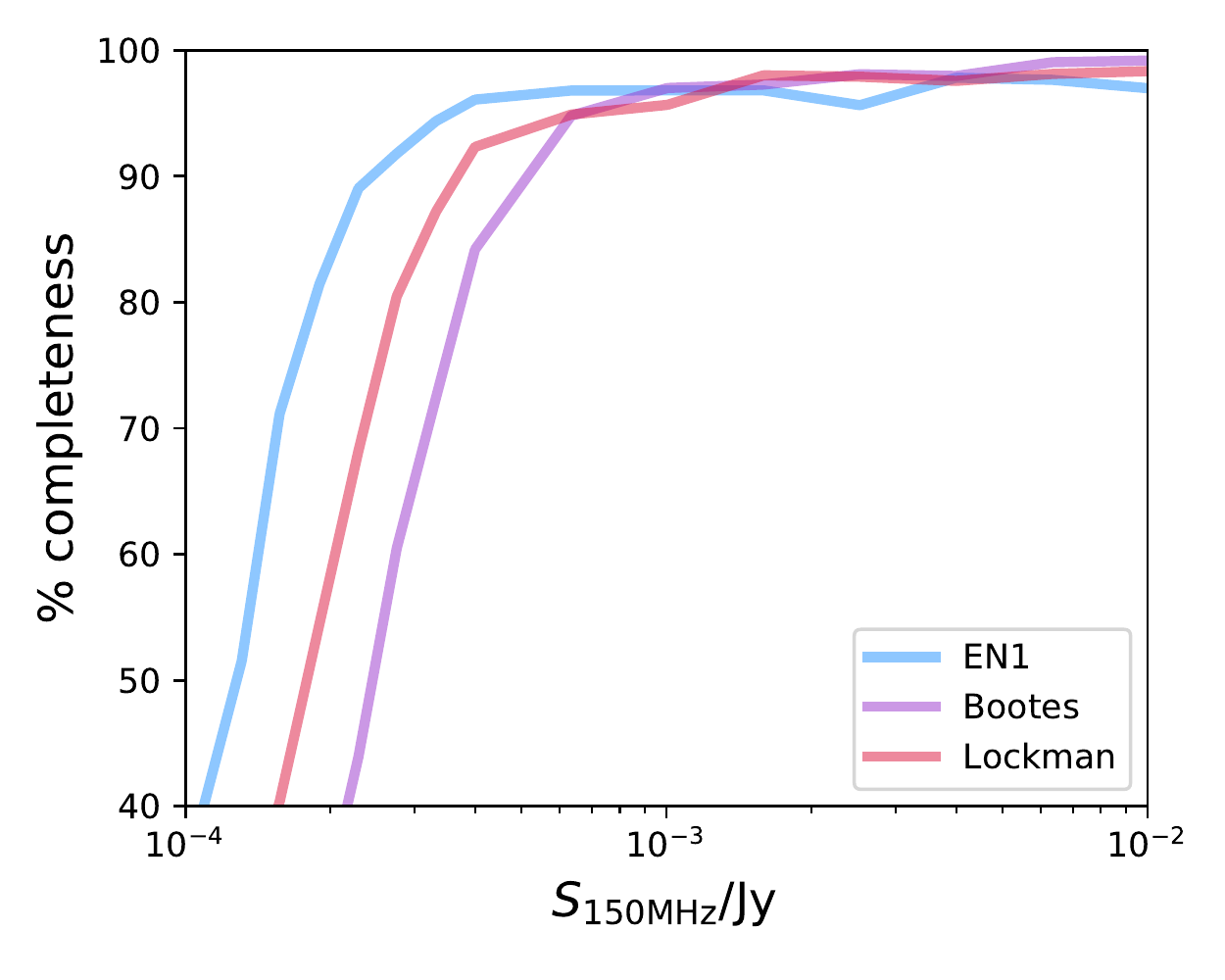}
\caption[]{Completeness as a function of $150\,\rm{MHz}$ flux density, for each of the three LoTSS Deep Fields. As described in Section \ref{sec:radio_completeness}, the completeness curves were calculated using source extraction of mock input sources from the radio images, using the same PyBDSF parameters as were used for the extraction of real sources. A realistic source size distribution was applied, based on the observed size distribution of star-forming galaxies with flux densities in the range $1-5\,\rm{mJy}$, where completeness is high. As expected based on total observing times, Elais-N1 has the deepest radio coverage, followed by the Lockman Hole, and then Bo{\"o}tes. Note that due to a very small number of injected mock sources overlapping with real sources and other injected sources, the completeness never quite reaches 100 per cent.}
\label{fig:completeness_sfg}
\end{figure}

\begin{figure}
\includegraphics[width=\columnwidth]{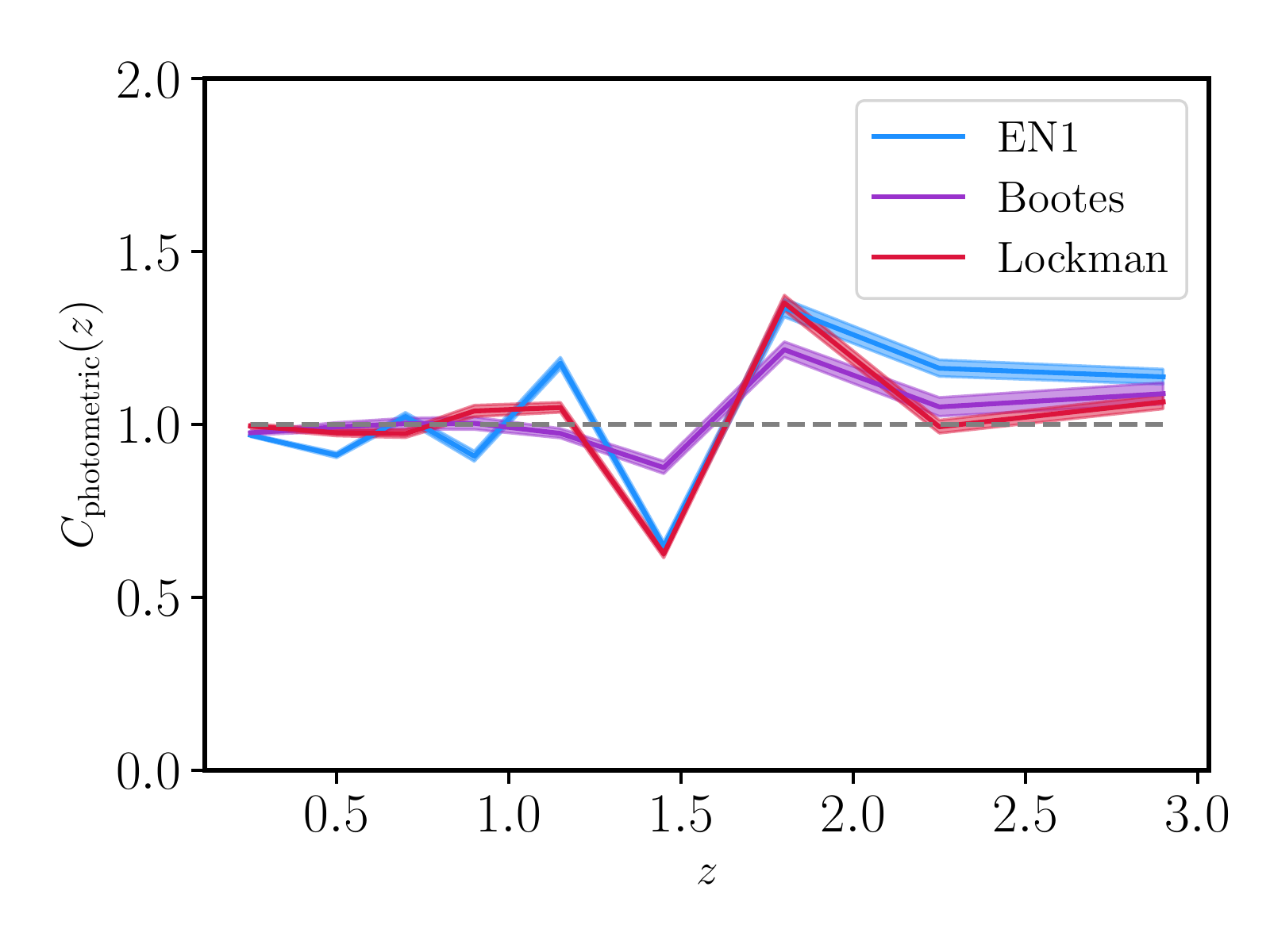}
\caption[]{Corrections applied to the number densities of radio sources derived in each redshift bin, for each field. To derive these corrections, a photometric redshift was drawn for each source from a flat redshift distribution between $z_{\rm{1,MIN}}$ and $z_{\rm{1,MAX}}$. The number of sources within each redshift bin studied was then compared to the number of sources within the bin that is derived using $z_{\rm{BEST}}$. The shaded regions are derived using $10,000$ bootstrapped samples. We apply these corrections to the number densities derived for each field, up to and including the $z=2.5-3.3$ bin. At higher redshifts, these corrections become less reliable due to long tails of photometric redshift probability distributions that peak at lower redshift. Hence, we do not apply corrections at $z>3.3$.}
\label{fig:photoz_corrections}
\end{figure}

\subsection{Radio completeness corrections}\label{sec:radio_completeness}
Completeness was calculated as a function of source flux density by simulating the source detection rate using the same techniques used to identify real sources. Approximately $100,000$ mock Gaussian sources of known angular extent and known $150\,\rm{MHz}$ total flux density were placed into regions of the LOFAR image that are covered by the optical imaging (accounting for masked regions). In practice, we insert $1,000$ mock sources at a time, a small number compared to the number of real sources in the LOFAR image, and repeat $
\sim1,000$ times; this is to avoid bias due to confusion. We used a continuous distribution of galaxy major axis size (from $6''$ to $30''$), and flux density (from $0.1\,\rm{mJy}$ to $40\,\rm{mJy}$). For each source, minor axis size was drawn from the distribution of minor axes of observed sources with roughly the same major axis size. Each mock source was separated from neighbours by at least twice its major axis size, so that sources did not overlap. PyBDSF was run on the image with inserted sources, using the same parameter choices as were adopted for the real radio source extraction. A source was deemed to be `recovered' if PyBDSF identified a source with flux density greater than $5\sigma$ of the local rms, within $2''$ of the position of the inserted source. The fraction of recovered to input sources was then calculated to characterise the completeness as a function of flux density and source size. To derive a single completeness curve per field, we then folded in our best estimate of the `true' source size distribution. This was derived from the size distribution of the star-forming galaxies in our sample where completeness reaches $\sim100$ per cent (source with flux densities in the range $1-5\,\rm{mJy}$). We are careful not to use only the very brightest sources, which will be biased in favour of nearby, spatially extended sources. Instead, we use the size distribution of sources at the point where the completeness curve flattens; in this way, we obtain the most similar size distribution to our sample, which is dominated by compact sources (note that intrinsically compact sources can have sizes larger than the beam due to the calibration). The size distribution of star-forming sources at $1-5\,\rm{mJy}$ is dominated by compact sources, with $84$ per cent having major axis sizes $<9''$ and a tail to larger values. We derive a single completeness curve, for each field (see Figure \ref{fig:completeness_sfg}) and provide these in Table \ref{Table:completeness}. The $50$ per cent completeness values are $128\,\mu\rm{Jy}$ for Elais-N1, $246\,\mu\rm{Jy}$ for Bo{\"o}tes, and $180\,\mu\rm{Jy}$ for the Lockman Hole. Equivalent curves were also derived assuming a source-size distribution appropriate for AGN, and are presented in \citep[][]{Kondapally2021a}. \\
\indent Radio completeness corrections were applied when constructing luminosity functions, as described in Section \ref{sec:details_lf}. We discard luminosity bins where applying the derived radio completeness correction results in a change of $>0.5\,\rm{dex}$.

\subsection{Incorporating uncertainties in photometric redshifts}\label{sec:photo_z_uncertainty}
When deriving luminosity functions, the redshift used for each source, $z_{\rm{BEST}}$, was the spectroscopic redshift where available, or otherwise the median of the primary photometric redshift solution. As discussed in Section \ref{sec:photzs}, the derivation of these redshifts combines template fitting and machine learning techniques, which yield particularly reliable results for the star-forming galaxies studied in this paper. However, as seen in Figure \ref{fig:demographics}, there is a dip in galaxy numbers at $z\sim1.5$, which may be driven by uncertainties in photometric redshifts. This effect is particularly marked in Elais-N1 and the Lockman Hole, which lack $H$-band data. The $H$-band is important in characterising the observed-frame wavelength of the $4000\angstrom$ break; for galaxies in the wavelength range $1.3<z<2.0$, the $4000\angstrom$ break falls between the $z$-band and the $J$-band. For these sources, $H$-band data constrains the flatter spectrum above the break, and hence enables more accurate redshifts to be derived. In the absence of the $H$-band, some photometric redshift aliasing can occur, and uncertainties on the derived photometric redshifts are larger. Here, we characterise the impact of these uncertainties on derived number densities in each of the three fields.\\ 
\indent Fully correcting for the uncertainties in photometric redshifts and their impact on the derived luminosity function would be complex. Assuming that the photometric redshift probability distribution of each source is robust, we could sample this finely and repeat the multi-code SED fitting, source classification and determination of a consensus SFR at each redshift for each source. When constructing the luminosity function, we could then draw bootstrapped samples, with the redshifts of each source drawn from its photometric redshift distribution. Given the substantial effort and computational expense required to derive source properties given just the best estimate redshift \citep[see][]{Best2021}, this would be impractical. There is also little evidence that the photometric redshift uncertainties are dependent on the radio luminosity of the source (which could require us to apply a luminosity-dependent correction). Instead we aim to take a simpler approach that will provide an approximate correction factor to the entire LF at each redshift for each field, $C_{\rm{photometric}}(z)$. This method accounts for aliasing in the photometric redshifts, under the assumption that there are no radio luminosity-dependent effects, and that source classifications and physical properties are not systematically changed by redshift errors.\\
\indent Our approach involves perturbing the redshift of each source according to its photometric redshift distribution, and calculating the change in numbers of galaxies that fall within each redshift bin studied. Instead of assuming $z_{\rm{TRUE}} = z_{\rm{BEST}}$, we draw a photometric redshift for each source from a flat redshift distribution between its $z_{\rm{1,MIN}}$ and $z_{\rm{1,MAX}}$. We then compare the new number of sources within each redshift bin to the number of sources within the bin that is derived using $z_{\rm{BEST}}$. We repeat this process $10,000$ times, using different random values between $z_{\rm{1,MIN}}$ and $z_{\rm{1,MAX}}$. This enables us to derive correction factors to the number densities of sources within each field and redshift range, as shown in Figure \ref{fig:photoz_corrections} and tabulated in Table \ref{Table:photoz_correction_table}. While most redshift bins have corrections $\sim1$ (i.e. no net gain or loss of sources, as the same number of sources are scattered into a given redshift bin as out of it), we see larger corrections required for Elais-N1 and the Lockman Hole (the fields that lack $H$-band data) at $z\sim1.5$. More sources get corrected into the $z\sim1.5$ redshift bin than out of it; this is because sources have $z_{\rm{BEST}}$ values that are pushed to higher/lower redshifts, but with large uncertainties that cover the $z\sim1.5$ bin. This leads to an upward correction to the luminosity function of $\sim50$ per cent for Elais-N1 and the Lockman Hole at $z\sim1.5$, with corresponding decreases for the neighbouring higher and lower redshift bins. We apply the derived corrections (as a multiplicative factor to $\phi$) up to and including the $z=2.5-3.3$ bin. At higher redshifts, these corrections become less reliable due to long tails of photometric redshift probability distributions that peak at lower redshift and are therefore not applied.

\begin{table}
\begin{center}
\begin{tabular}{c|c|c}
\hline
$\log_{10}(L_{150\,\rm{MHz}}/\rm{W}\,Hz^{-1})$ & $N_{\rm{sources}}$ & $\log_{10}(\phi/\rm{Mpc}^{-3}\log_{10}{L}^{-1})$ \\ 
\hline
$20.75$ & $6$ & $-2.22^{+0.18}_{-0.31}$ \vspace{0.1cm}\\
$21.05$ & $20$ & $-2.33^{+0.11}_{-0.13}$ \vspace{0.1cm}\\
$21.35$ & $84$ & $-2.25^{+0.07}_{-0.08}$ \vspace{0.1cm}\\
$21.65$ & $171$ & $-2.42^{+0.06}_{-0.07}$ \vspace{0.1cm}\\
$21.95$ & $397$ & $-2.45\pm0.05$ \vspace{0.05cm}\\
$22.25$ & $888$ & $-2.50\pm0.05$ \vspace{0.05cm}\\
$22.55$ & $1562$ & $-2.61\pm0.04$ \vspace{0.05cm}\\
$22.85$ & $1579$ & $-2.81\pm0.02$  \vspace{0.05cm}\\
$23.15$ & $791$ & $-3.16\pm0.02$ \vspace{0.05cm}\\
$23.45$ & $257$ & $-3.66\pm0.03$ \vspace{0.05cm}\\
$23.75$ & $49$ & $-4.39^{+0.06}_{-0.07}$ \vspace{0.1cm}\\
$24.05$ & $8$ & $-5.18^{+0.14}_{-0.20}$ \vspace{0.1cm}\\
$24.35$ & $2$ & $-5.78\pm0.30$ \vspace{0.05cm}\\
$24.65$ & $1$ & $-6.09\pm0.30$
\end{tabular}
\caption{The local ($0.03<z<0.30$) $150\,\rm{MHz}$ luminosity function for galaxies with radio emission dominated by star formation, as shown in Figure \ref{fig:parametric_fit}. Equivalent data for radio-loud AGN are presented in \protect\cite{Kondapally2021a}.}
\label{Table:lf_at_z0}
\end{center}
\end{table}

\begin{figure*}
\includegraphics[width=8.1cm]{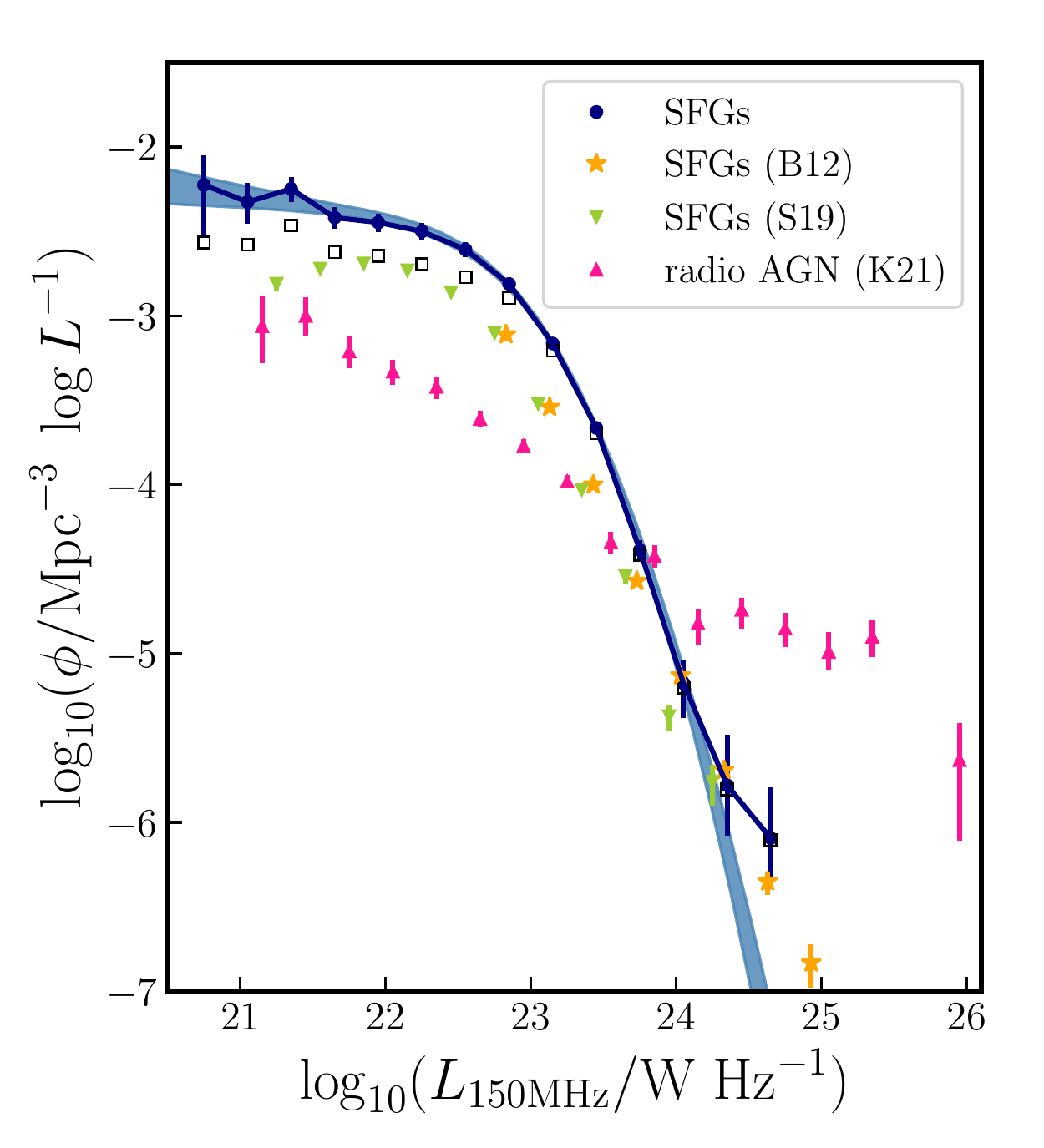}
\includegraphics[width=9.6cm]{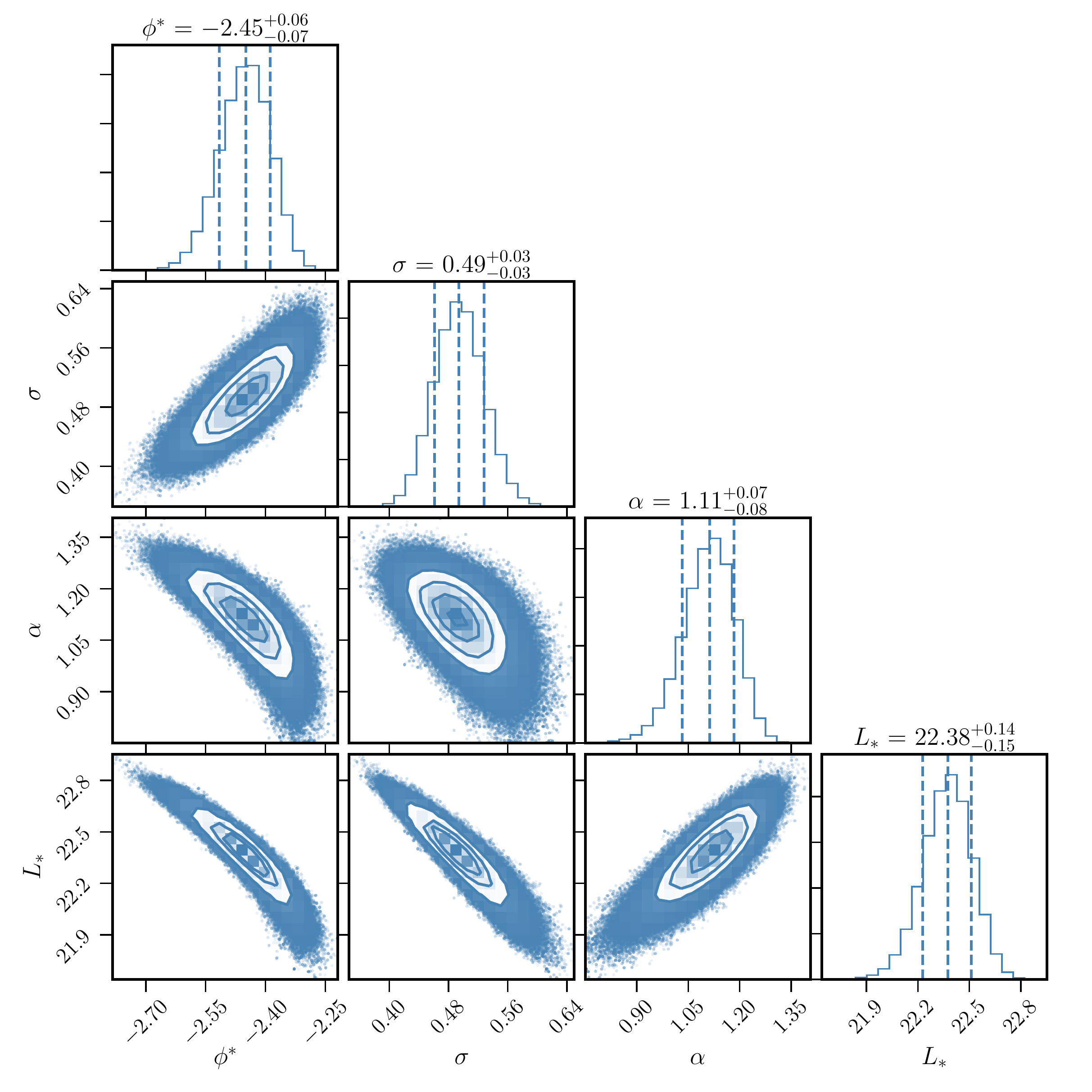}
\caption[]{Characterisation of the form of the low redshift $150\,\rm{MHz}$ luminosity function for star-forming galaxies. Left: the luminosity function of star-forming galaxies in the redshift range $z=0.03-0.30$, constructed using sources in the three LOFAR Deep Fields that show no evidence of AGN-driven radio emission. The dark blue points show our estimate including the radio completeness corrections, with error bars calculated using $1000$ bootstrapped samples. Data points without the completeness correction are shown by open squares (black). Our data are fitted with a \cite{Saunders1990} parametrisation; the pale blue shaded region shows the $16^{\rm{th}}-84^{\rm{th}}$ percentile of the posterior distribution of one example of the binning scheme. For comparison, the local ($z<0.3$) luminosity function of $150\,\rm{MHz}$-selected star-forming galaxies from LoTSS DR1, derived by \cite{Sabater2019}, is shown in green. The local ($z<0.3$) luminosity function of $1.4\,\rm{GHz}$-detected star-forming galaxies (scaled using a radio spectral index $\alpha=-0.7$), derived by \cite{Best2012}, is shown in orange. The $z=0.03-0.30$ luminosity function for radio-selected AGN, derived by \cite{Kondapally2021a}, is shown in pink. Right: the posterior distributions of the four fitted parameters, which are well-constrained. The fitting was repeated for 18 combinations of luminosity bin size and position, and the results were then averaged. The fitted values depend little on the choice of luminosity binning. Shown in the right hand panel is just one example of the posteriors from one choice of binning; we use the average fitted values of $\sigma=0.49$ and $\alpha=1.12$ in our fits at other redshifts.}
\label{fig:parametric_fit}
\end{figure*}

\begin{table*}
\begin{center}
\begin{tabular}{l|c|c|c|c}
\hline
Redshift & $N_{\rm{sources}}$ & $\log_{10}(L_{\star,150\rm{MHz}}/\rm{W}\,Hz^{-1})$ & $\log_{10}(\phi_{\star}/\rm{Mpc}^{-3}\log{L}^{-1})$ & $\rm{SFRD}/M_{\odot}\,\rm{yr}^{-1}\,\rm{Mpc}^{-3}$ \\ 
\hline
$0.1-0.4$ & $8510$ & $22.52\pm0.02$ & $-2.58\pm0.03$ & $0.023\pm0.001$\\
$0.4-0.6$ & $6908$ & $22.78\pm0.03$ & $-2.59\pm0.06$ & $0.040\pm0.004$\\
$0.6-0.8$ & $6788$ & $22.97\pm0.03$ & $-2.60\pm0.07$ & $0.059\pm0.006$\\
$0.8-1.0$ & $5214$ & $23.28\pm0.04$ & $-2.95\pm0.07$ & $0.053\pm0.005$\\
$1.0-1.3$ & $6957$ & $23.46\pm0.03$ & $-2.99\pm0.07$ & $0.069\pm0.007$\\
$1.3-1.6^{*}$ & $2897$ & $23.72\pm0.07$ & $-2.88\pm0.15$ & $0.079\pm0.018$\\
$1.6-2.0^{*}$ & $6573$ & $23.94\pm0.03$ & $-3.37\pm0.06$ & $0.111\pm0.017$\\
$2.0-2.5$ & $4516$ & $24.08\pm0.03$ & $-3.45\pm0.06$ & $0.096\pm0.010$\\
$2.5-3.3$ & $4101$ & $24.28\pm0.04$ & $-3.65\pm0.08$ & $0.092\pm0.011$\\
$3.3-4.6$ & $1951$ & $23.79\pm0.04$ & $-4.50\pm0.07$ & $0.039\pm0.004$\\
\end{tabular}
\caption{Parameters derived from fits of the \protect\cite{Saunders1990} luminosity function to the data shown in Figure \ref{fig:all_rq_lfs}. For all redshift bins apart from $z=1.3-1.6$, the luminosity function is derived using data from all three LOFAR Deep Fields. At $z=1.3-2.0$, our best estimates of the luminosity function are from the Bo{\"o}tes field alone, due to the larger photometric uncertainties in Elais-N1 and the Lockman Hole (see Section \ref{sec:photo_z_uncertainty}). $\phi_{\star}$ represents our best estimate of the characteristic number density, after correcting for photometric redshift uncertainties, as described in Section \ref{sec:photo_z_uncertainty}. We also provide our estimate of the cosmic star formation rate density integrated down to $0.03\,L_{\star}$ at each redshift, calculated using a $L_{150\,\rm{MHz}}-\rm{SFR}$ calibration derived from the same dataset, in combination with a correction for the scatter in this relation (see Section \ref{sec:L_sfr_cal}). SFRD estimates for each individual field are provided in Table \ref{Table:sfrd_each_field}.}
\label{Table:lf_with_z}
\end{center}
\end{table*}

\begin{figure*}
\includegraphics[width=16.6cm]{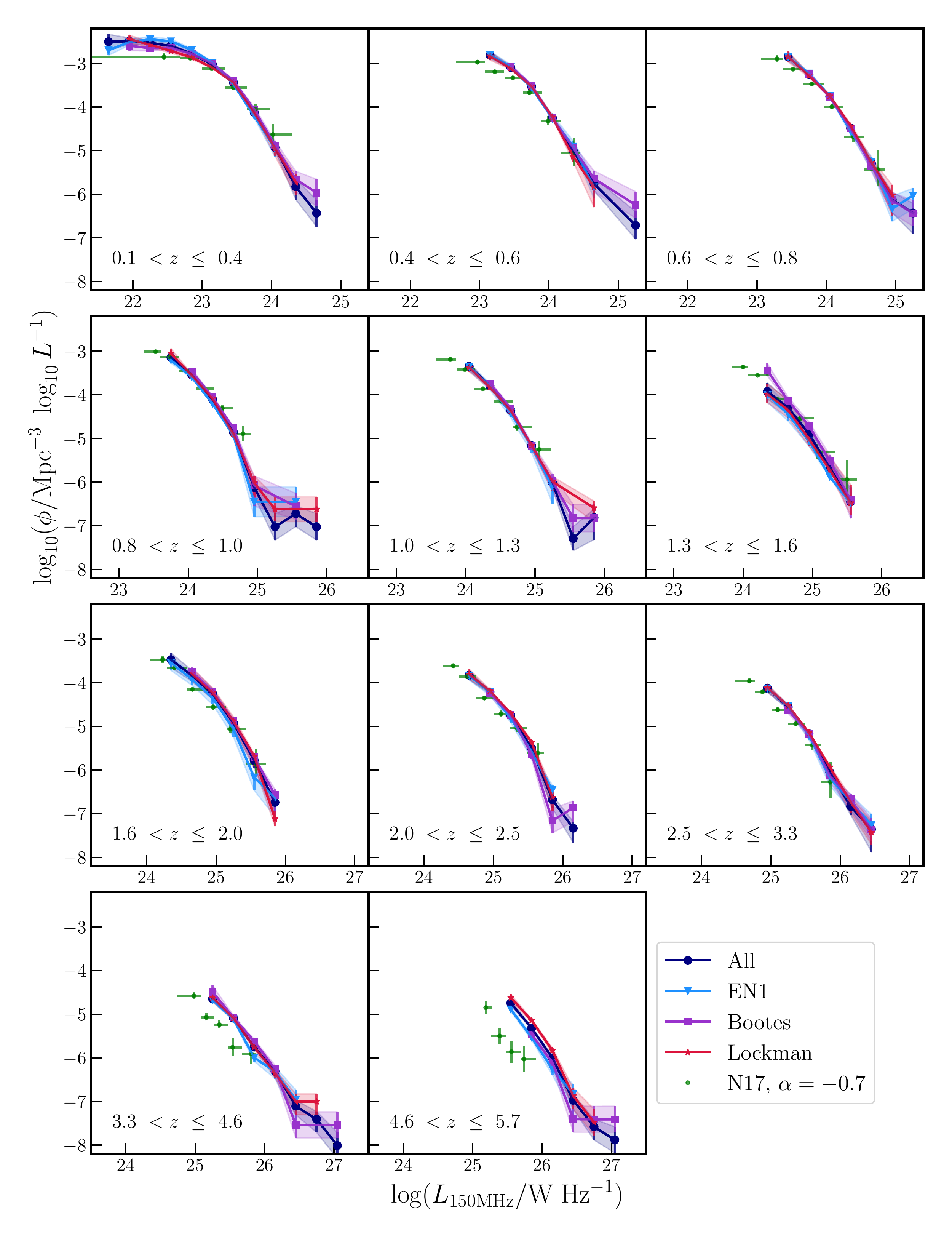}
\vspace{-0.5cm}
\caption[]{The $150\,\rm{MHz}$ luminosity function for galaxies with radio emission dominated by star formation, for the three LOFAR Deep Fields: Elais-N1 (mid blue), Bo{\"o}tes (purple) and the Lockman Hole (red), as well as for the combined sample (navy). Note the change in x-axis scale for the different redshifts. Corrections for uncertainties in photometric redshifts have been applied as a fixed scaling for a given redshift for each field, as described in Section \ref{sec:photo_z_uncertainty}. Corrections for radio completeness have also been made; data are only plotted for bins where this correction is $<0.5\,\rm{dex}$. The luminosity functions show excellent consistency between the three fields and also (except at the highest redshifts) good agreement with the estimates of \cite{Novak2017}, shown in green, which have been scaled from $3\,\rm{GHz}$ using a radio spectral index $\alpha=-0.7$.}
\label{fig:all_rq_lfs}
\end{figure*}

\subsection{Parametric fits to the radio luminosity function}
\begin{figure}
\includegraphics[width=\columnwidth]{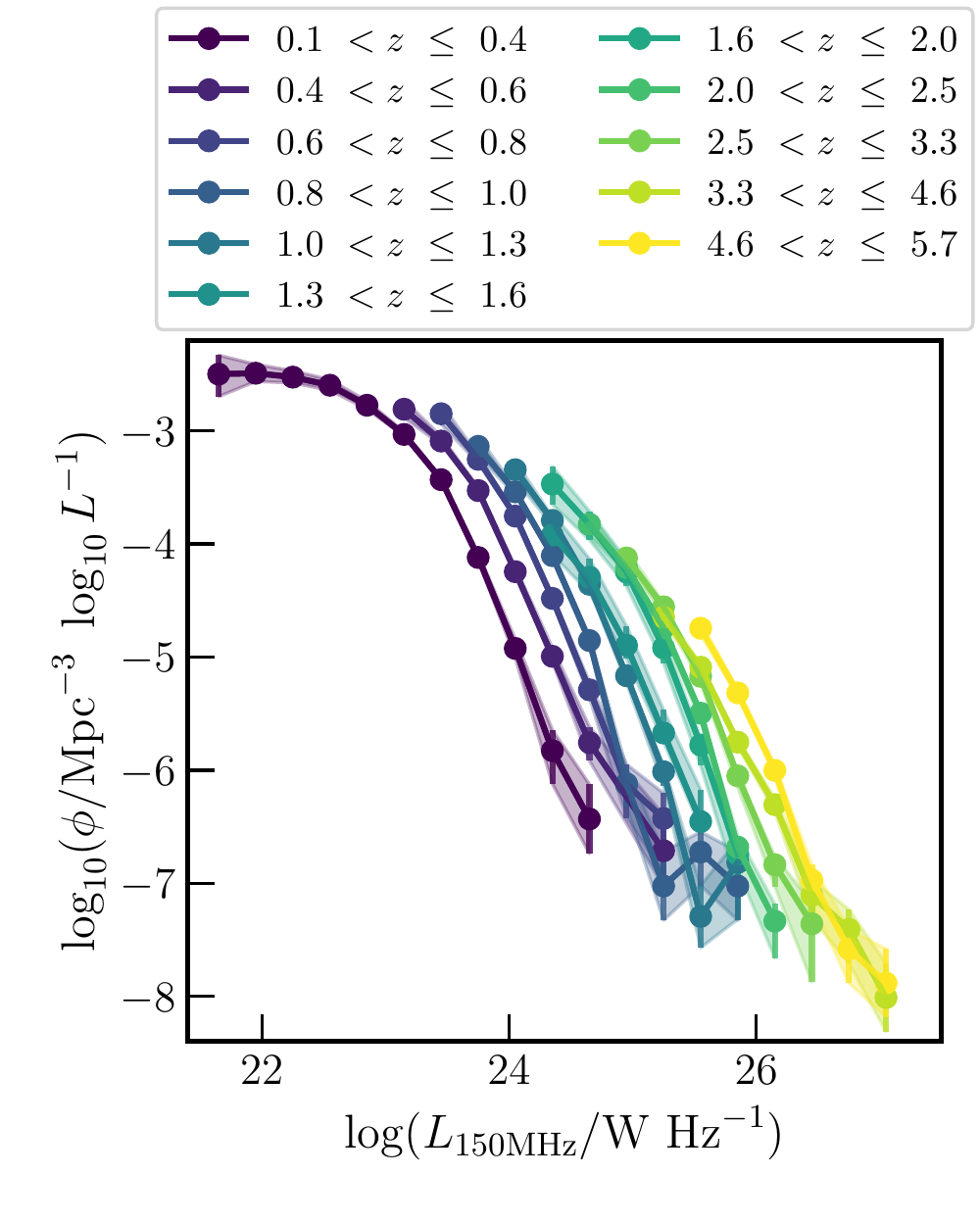}
\vspace{-0.5cm}
\caption[]{Radio luminosity functions derived using all three LOFAR Deep Fields and presented in Figure \ref{fig:all_rq_lfs}, plotted on a single figure to illustrate the strong redshift evolution.}
\label{fig:oneplot_rq_lfs}
\end{figure}

To characterise the evolution of the $150\,\rm{MHz}$ luminosity function, we fit each of the derived LFs with a parameterised expression. Numerous studies of radio luminosity functions of star-forming galaxies \citep{Saunders1990,Best2005,Novak2017} have adopted the following parametrisation, first used by \cite{Sandage1979}:
\begin{equation}\label{eq:lf_fit}
    \phi(L) = \phi_{\star}\Bigg(\frac{L}{L_{\star}}\Bigg)^{1-\alpha}\exp\Bigg[-\frac{1}{2\sigma^{2}}\log^{2}\Bigg(1+\frac{L}{L_{\star}}\Bigg)\Bigg],
\end{equation}
where $\phi_{\star}$ provides the normalisation, $L_{\star}$ is the luminosity at the turnover, $\alpha$ is the faint end slope, and $\sigma$ describes the steepness at the bright end. \\
\indent To enable fits to the radio luminosity functions of high redshift galaxies in the COSMOS-VLA survey, \cite{Novak2017} refitted previously-derived local $1.4\,\rm{GHz}$ radio luminosity functions of star-forming galaxies. They derived the best-fit parameters: $\log_{10}(\phi_{\star,\,1.4\rm{GHz}}/\rm{Mpc^{-3}\,dex^{-1}})=-2.45$, $\log_{10}(L_{\star,\,1.4\rm{GHz}}/\rm{W\,Hz^{-1}})=21.27$, $\alpha=1.22$, and $\sigma=0.3$. They then fixed $\alpha$, and $\sigma$ in their fits at higher redshifts. \\
\indent We repeat this process to derive $\alpha$ and $\sigma$ self-consistently from our own LOFAR Deep Fields data. We fit the low redshift ($z=0.03-0.3$) luminosity function, as shown in Figure \ref{fig:parametric_fit}. We exclude the first luminosity bin from the fit due to the larger ($>0.3\,\rm{dex}$) completeness corrections for the faintest sources. We also exclude the final luminosity bin due to the increased potential importance of misclassification for the brightest sources; as shown in Figure \ref{fig:parametric_fit}, radio AGN dominate the whole radio sample here. We repeat this fitting process for 18 combinations of luminosity bin size and position, and average over the results to derive the following best fit parameters: $\log_{10}(\phi_{\star}/\,\rm{Mpc^{-3}\,dex^{-1}})=-2.46\pm0.01$, $\sigma=0.49\pm0.01$, $\alpha=1.12\pm0.01$ and $\log_{10}(L_{\star}/\mathrm{W\,Hz}^{-1})=22.40^{+0.02}_{-0.03}$. We fix $\sigma=0.49$ and $\alpha=1.12$ for the remainder of this work.\\
\indent We overplot several measurements from the literature in Figure \ref{fig:parametric_fit}. \cite{Kondapally2021a} derived radio luminosity functions of radio-selected AGN in the three LoTSS Deep Fields as a function of redshift. Their $z=0.03-0.3$ measurement is plotted here. We note the very different shapes of the luminosity functions for the two populations. As expected, AGN dominate the source counts at the highest luminosities. At $L_{150\,\rm{MHz}}\sim10^{23.75}\,\rm{W\,Hz^{-1}}$, there are approximately equal contributions of AGN and SFGs to the source counts. We also show luminosity functions constructed for the star-forming galaxy population by \cite{Best2012} and \cite{Sabater2019}. While in good agreement with each other, these data lie below ours at all but the brightest luminosities. Partly, this is due to the lack of radio completeness corrections in the earlier work. Differing redshift distributions of the samples will also contribute: for our sample, the median redshift of sources included in the $z=0.03-0.3$ subsample is $z_{\rm{median}}=0.20$. The median redshift of star-forming sources using in both \cite{Best2012} and \cite{Sabater2019} will be lower, due to less sensitive radio imaging. Coupled with the strong redshift evolution seen for these samples (see Figure \ref{fig:oneplot_rq_lfs}), this will naturally lead to a small offset in derived luminosity functions.

\begin{figure*}
\includegraphics[scale=0.435]{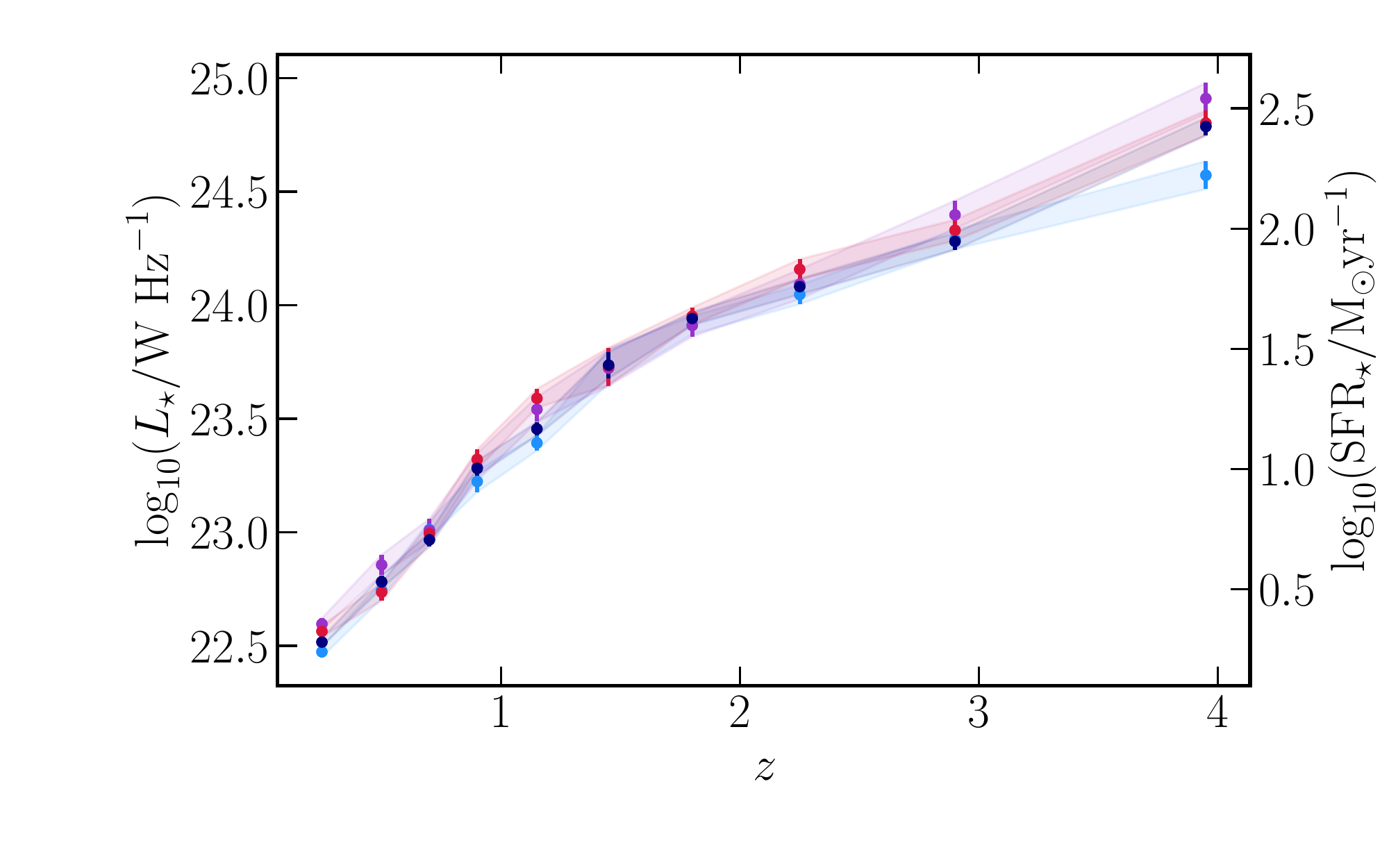}
\includegraphics[scale=0.435]{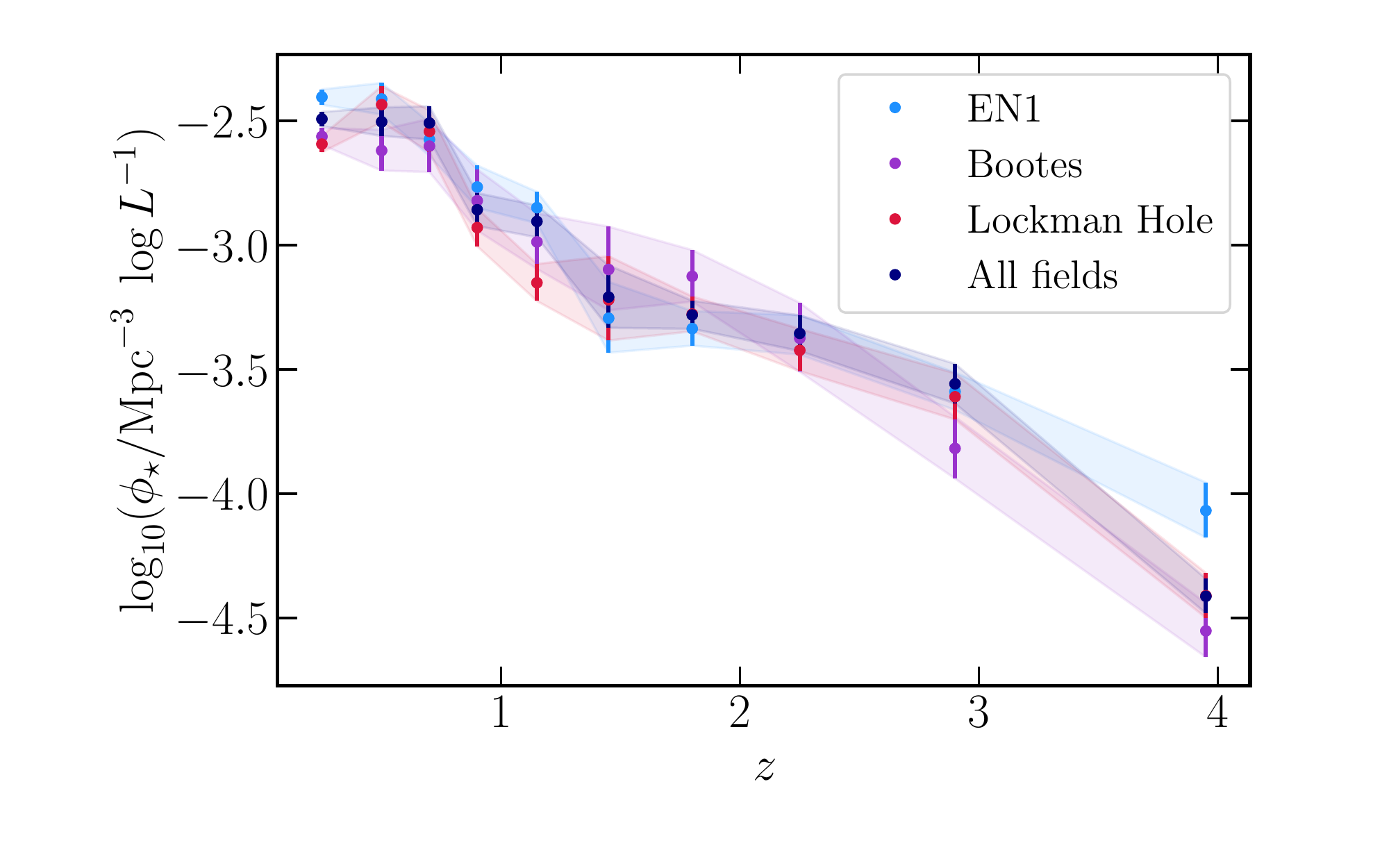}
\caption[]{The evolution of best-fitting luminosity function parameters, for the three LoTSS Deep Fields individually and combined. Values of $\phi_{\star}$ have been corrected for uncertainties in photometric redshifts, as described in Section \ref{sec:photo_z_uncertainty}. To guide more physical intuition, we show in the left-hand panel the characteristic star formation rate corresponding to the characteristic luminosity, derived using the $L_{150\,\rm{MHz}}-\rm{SFR}$ relation given in Equation \ref{eq:Dan_relation}. The derived parameter values are generally in good agreement across the fields. The three estimates of $\phi_{\star}$ and $L_{\star}$ diverge at the highest redshift ($z\sim4$), where $L_{\star}$ falls well below the sensitivity limits of the radio data.}
\label{fig:Lstar_phistar_evo}
\end{figure*}

\subsection{Evolution of the $\mathbf{150\,\rm\bf{MHz}}$ luminosity function from $\mathbf{z=0}$ to $\mathbf{z=5}$}\label{sec:results_lf}
The luminosity functions derived for each of the LOFAR Deep Fields, as well as for the combination of the three, are shown in Figure \ref{fig:all_rq_lfs}. These luminosity functions have been corrected for radio completeness and uncertainties in photometric redshifts. The fields display excellent agreement with each other, indicating that cosmic variance effects are minimal (as expected over such large areas). As a sanity check, we confirm that our estimates agree well with data from \cite{Bonato2021}, who measured luminosity functions of star-forming galaxies and radio quiet AGN out to $z=2.8$ using Deep Fields data in the Lockman Hole. Their measurements are consistent with modelled populations in the Tiered Radio Extragalactic Continuum Simulation (T-RECS; \citealt{Bonaldi2019}). As shown in Figure \ref{fig:all_rq_lfs}, we also find very good agreement with the luminosity functions derived for VLA-COSMOS by \cite{Novak2017}, once luminosities are scaled from $3\,\rm{GHz}$, under the assumption of a fixed radio spectral index (we use $\alpha=-0.7$), apart from at the very highest redshifts studied, where number densities in COSMOS drop more significantly. In Figure \ref{fig:oneplot_rq_lfs}, we show the derived luminosity functions at all redshifts on a single panel, for easier comparison.\\
\indent At each redshift, we fit the radio luminosity function for each of the three fields, as well as for all fields combined, with the parametrisation presented in Equation \ref{eq:lf_fit}, fixing $\sigma=0.49$ and $\alpha=1.12$. The inferred $L_{\star}$ and $\phi_{\star}$ values (displayed in Figure \ref{fig:Lstar_phistar_evo} and Table \ref{Table:lf_with_z}) are in good agreement between the three fields, except at the very highest redshifts, where $L_{\star}$ falls well below the sensitivity limits of the radio data. $L_{\star}$ increases monotonically out to at least $z\sim3$, displaying an evolution of $>1.5\,\rm{dex}$ between $z\sim0.25$ and $z\sim3$. $\phi_{\star}$ remains roughly constant back to $z\sim0.8$ but then falls steeply at higher redshifts, decreasing by $>1\,\rm{dex}$ between $z\sim0.7$ and $z\sim3$.

\section{Reliable conversion of $L_{150\rm{MHz}}$ to SFR}\label{sec:L_sfr_calibration}
\subsection{The star formation rate function constructed using different SFR estimates}\label{sec:sfrf_various}
\begin{figure*}
\includegraphics[width=16cm]{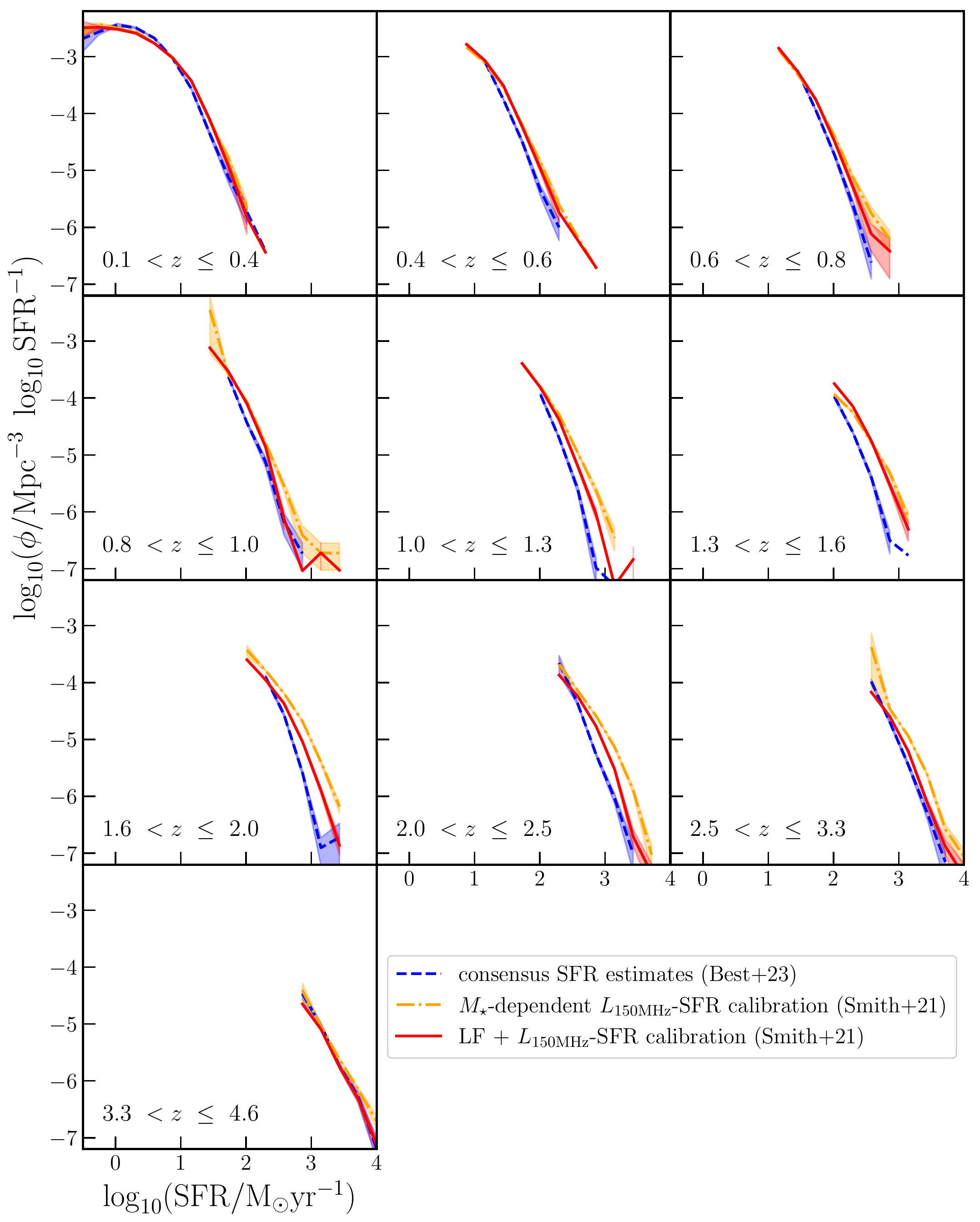}
\caption[]{The SFR function derived from the $150\,\rm{MHz}$-selected sample in all three LOFAR Deep Fields, constructed using different SFR estimates. The SFR function derived using the consensus SFR estimates derived by \cite{Best2021} is shown in  dashed blue. The solid red line shows that derived using the $L_{150\,\rm{MHz}}-\rm{SFR}$ conversion derived by \cite{Smith2021}. The dashed yellow line shows that derived using the mass-dependent $L_{150\,\rm{MHz}}-\rm{SFR}$ conversion derived by \cite{Smith2021}. Despite the $L_{150\,\rm{MHz}}-\rm{SFR}$ conversion being derived from the same LOFAR Deep Fields data, the derived SFR functions differ above $\rm{SFR}_{\star}$, in some cases by an order of magnitude.} 
\label{fig:L-SFR_cal_diffs}
\end{figure*}

In this section, we derive the star formation rate function (SFRF) using multiple SFR estimates. In principle, we can transform the radio LFs to SFRFs using a previously-calibrated relation between $150\,\rm{MHz}$ radio luminosity and SFR \citep{Rivera2017,Gurkan2018,Smith2021}. \cite{Smith2021} derived the following relation using LOFAR data in EN1, using {\small MAGPHYS}-derived SFRs for radio and NIR-selected sources at $z<1$:
\begin{equation}\label{eq:Dan_relation}
\begin{split}
   \log_{10}(L_{150\rm{MHz}}/\rm{W}\,Hz^{-1}) = (22.221 \pm 0.008) + \\ (1.058 \pm 0.007) \log_{10}(\rm{SFR}/M_{\odot}\rm{yr}^{-1}).
\end{split}
\end{equation}
A consistent relation was derived using the ‘ridgeline’ approach of \cite{Best2021}, and we confirm that re-fitting SFR as a function of $L_{150\rm{MHz}}$ (as opposed to $L_{150\rm{MHz}}$ as a function of SFR) gives a similar result. We therefore use the \cite{Smith2021} relation to derive SFR functions, essentially scaled luminosity functions; these are plotted in red in Figure \ref{fig:L-SFR_cal_diffs}. Although the uncertainties on the best-fitting parameters are small, \cite{Smith2021} showed that there was substantial intrinsic scatter on the relation ($\sim0.3\,\rm{dex}$) at $\rm{SFR}=10-100\,M_{\odot}\rm{yr}^{-1}$. \\
\indent Adding an additional parameter such as stellar mass can decrease the scatter on this relation. Using shallower LOFAR data in the {\it{Herschel}}-ATLAS NGP field, and focusing on star-forming galaxies at $z<0.4$, \cite{Gurkan2018} found evidence of a dependence of the radio luminosity on both SFR (the primary driver) and stellar mass (a secondary parameter). They derived a mass-dependent $L_{150\,\rm{MHz}}-\rm{SFR}$ relation, with a break in the relation around $\rm{SFR}=1\,\rm{M_{\odot}}\rm{yr}^{-1}$. They speculated that this may be due to alternative mechanisms for generating cosmic rays in the lowest mass galaxies (see \citealt{Schober2022} for further discussion of the physical cause of the stellar mass dependence of the infrared-radio correlation). \cite{Smith2021} built on this work by re-deriving a stellar mass-dependent relation in EN1 using the LOFAR Deep Fields data used in this paper:
\begin{equation}\label{eq:Dan_relation_mass}
\begin{split}
   \log_{10}(L_{150\rm{MHz}}/\rm{W}\,Hz^{-1}) = (22.218 \pm 0.016) + \\ (0.903 \pm 0.012) \log_{10}(\rm{SFR}/M_{\odot}\rm{yr}^{-1}) + \\ (0.332 \pm 0.037) \log_{10}(\rm{M_{\star}}/10^{10}\,M_{\odot}).
\end{split}
\end{equation}
They argued that the stellar mass dependence of the relation can introduce substantial systematic errors (of order $0.5\,\rm{dex}$) on SFRs derived from $L_{150\rm{MHz}}$ alone, particularly in cases where the sample for which SFRs are derived has a different stellar mass distribution to the sample from which the relation was derived. They noted that these offsets are potentially larger than the intrinsic scatter in the $L_{150\,\rm{MHz}}-\rm{SFR}$ relation.\\
\indent We construct an SFR function using SFR estimates derived from both $L_{150\rm{MHz}}$ and stellar mass as follows. For each galaxy, we input the `consensus' stellar mass estimate provided by \cite{Best2021} and the radio luminosity into Equation \ref{eq:Dan_relation_mass} to obtain a new SFR estimate. We then construct the SFR function in a similar way to the luminosity function, applying radio luminosity and photometric uncertainty corrections on a source-by-source basis. We overlay the SFR function derived for each redshift bin in yellow in Figure \ref{fig:L-SFR_cal_diffs}. The SFR functions diverge from those derived from radio luminosity without stellar mass only at very high SFRs and at redshifts beyond the range that was used to derive the $L_{150\,\rm{MHz}}-\rm{SFR}$ relation. \\
\indent We also construct SFR functions using the consensus SFR estimates presented by \cite[][blue lines]{Best2021}; as described in Section \ref{sec:sfrs_masses}, these were derived using SED fits to the multi-wavelength data, rather than a single wavelength indicator. We compare the SFRFs derived from the two radio luminosity calibrations (red and yellow lines) to those derived using consensus SFR estimates (blue lines). As seen most clearly at low redshifts, below the break of the function ($\rm{SFR}_{\star}$), the SFRF estimates are in reasonable agreement. However, they can differ substantially at the highest SFRs, in some cases by an order of magnitude. This is despite the $L_{150\,\rm{MHz}}-\rm{SFR}$ conversion also being derived from the LOFAR Deep Fields data and SED fits. At first glance, it is worrying that we obtain such different SFRFs when using SFRs derived using different methods, given that the parent radio samples used are the same. The difference is largest at high SFRs, so is of particular concern for studies like ours, where most of the SFRF data points are above $\sim\rm{SFR}_{\star}$ at all but the lowest redshifts studied. Since the star formation rate density is derived by integrating the SFR function, this could have implications for the normalisation and the shape of the inferred cosmic star formation rate density-redshift relation. In Section \ref{sec:L_sfr_cal}, we show that this effect arises due to scatter in the $L_{150\,\rm{MHz}}-\rm{SFR}$ relation, which we have not accounted for thus far, and we develop a method to correct for this bias.

\begin{figure*}
\includegraphics[width=8.7cm]{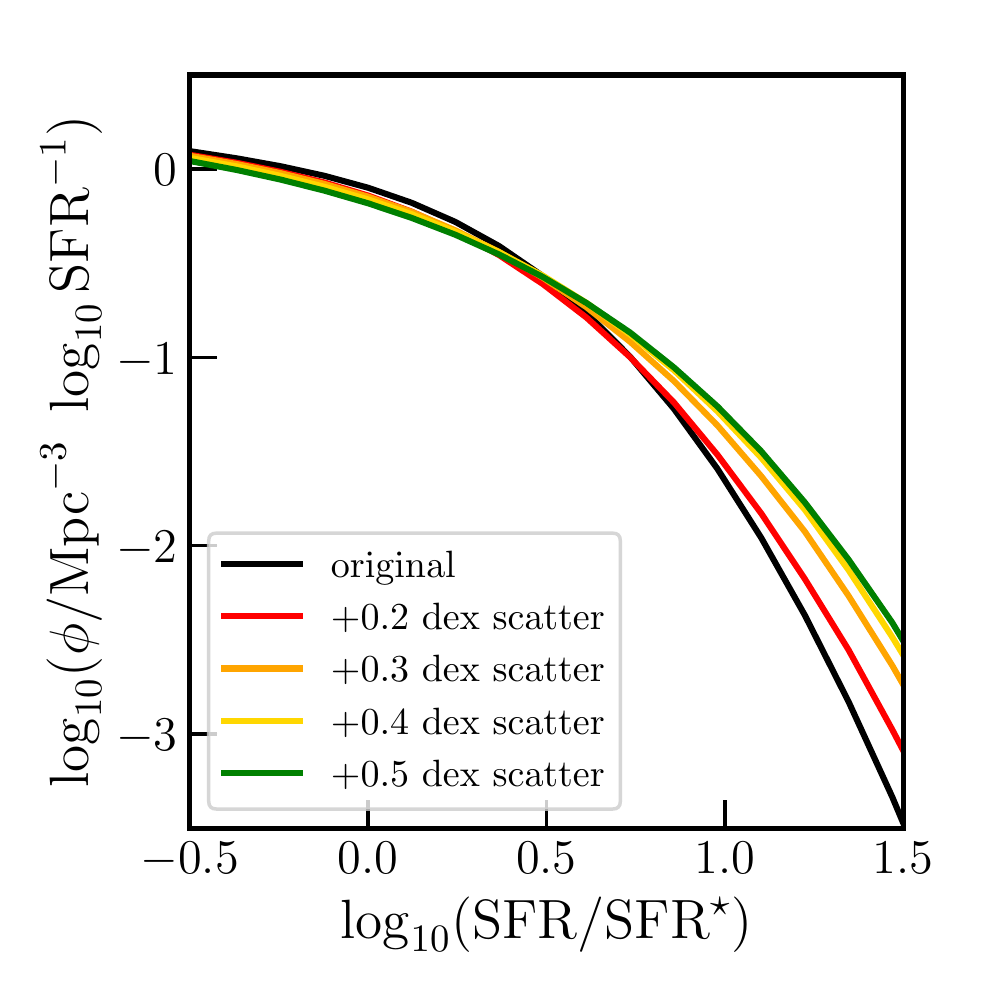}
\includegraphics[width=8.7cm]{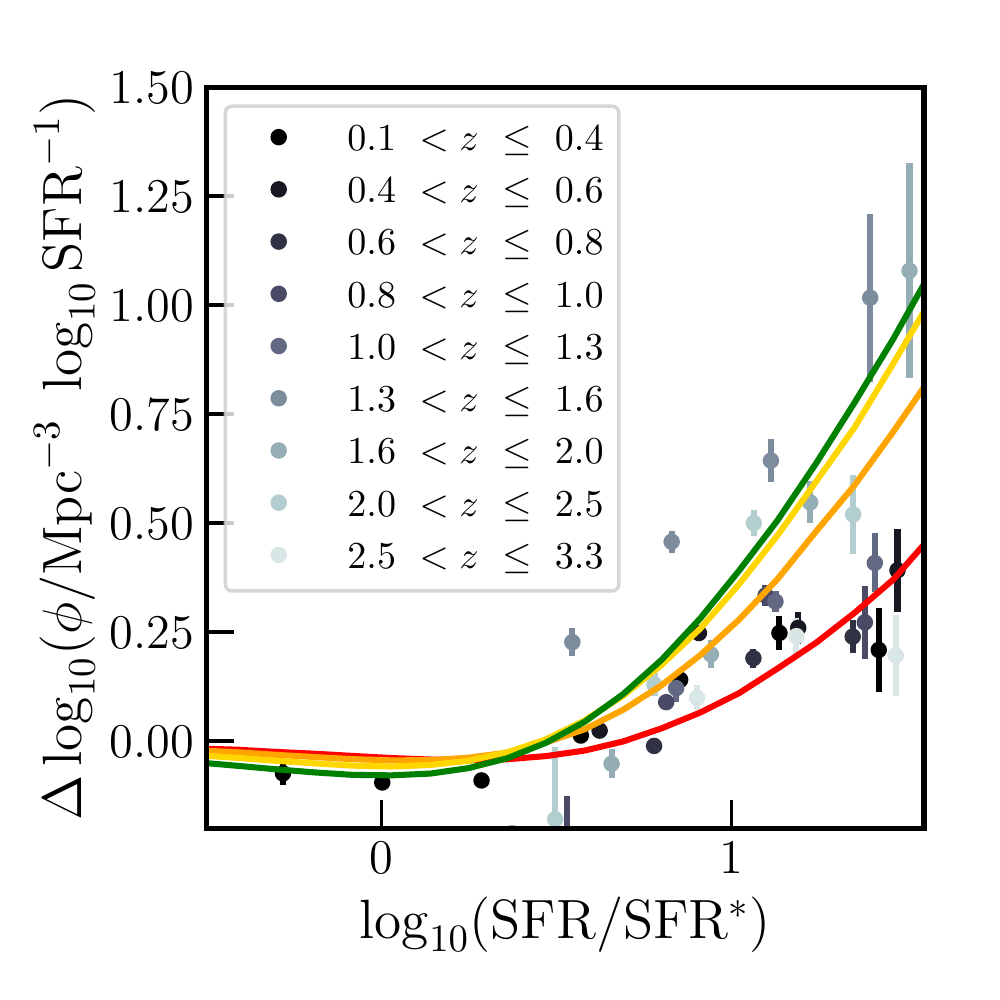}
\caption[]{Constraining the scatter in the $L_{150\,\rm{MHz}}-\rm{SFR}$ relation. Left: simulated deviation of the star formation rate function, as measured from $L_{150\,\rm{MHz}}$, from the input star formation rate function (black), for different values of the scatter in the $L_{150\,\rm{MHz}}-\rm{SFR}$ relation. Larger scatter in the relation causes the observed number density of galaxies to increase at the bright end, causing a gentler fall-off of the exponential. This leads to an overestimation of the cosmic star formation rate density. Right: the method used to constrain the scatter in the $L_{150\,\rm{MHz}}-\rm{SFR}$ relation. Grey data points show offsets between the star formation rate function constructed using consensus SFR estimates, and that constructed by applying a single $L_{150\,\rm{MHz}}-\rm{SFR}$ relation to the $150\,\rm{MHz}$ luminosity function. Coloured lines show the modelled offsets for different values of scatter in the relation. From this, we calibrate the scatter to be $\sim0.2-0.4\,\rm{dex}$. Integrating these coloured lines, and comparing to the integrated black line (no scatter), we obtain scaling factors to correct values of the cosmic star formation rate density derived using a single $L_{150\,\rm{MHz}}-\rm{SFR}$ conversion. These are: $\rm{CORR_{SFRD}} = [1.0,0.96,0.93]$, for scatter $\Sigma=[0.2,0.3,0.4]\,\rm{dex}$.}
\label{fig:L-SFR_cal}
\end{figure*}

\subsection{The impact of the calibration between radio luminosity and SFR}\label{sec:L_sfr_cal}
As discussed in Section \ref{sec:sfrf_various}, the derived star formation rate function depends strongly on the method used to estimate star formation rates. Above $\sim\rm{SFR}_{\star}$, the SFR function is highly dependent on the method used to infer SFRs. We expect that this is, at least in part, due to a combination of Eddington bias \citep{Eddington1913} and the scatter in the $\rm{L_{150\,MHz}}-\rm{SFR}$ relation. In this section, we explore the effects of different amounts of scatter on the derived star formation rate function. Using a simple simulation, we demonstrate the magnitude of the bias and derive correction factors. \\
\indent We begin by generating $\sim300$ million mock sources with star formation rates drawn from a \cite{Saunders1990} function with default values of $\rm{SFR_{\star}=0}$ and $\phi_{\star}=0$. We set $\sigma=0.38$ (the average value of $\sigma$ from fits to the consensus estimate-derived SFR function at $z\lesssim1$). The modelled base SFR function is shown in black in the left-hand panel of Figure \ref{fig:L-SFR_cal}. We then simulate the radio luminosities of the mock sources using Equation \ref{eq:Dan_relation}, adding scatter drawn from a Gaussian distribution with $\Sigma=[0.2,0.3,0.4,0.5]\,\rm{dex}$, but truncated at $0.7\,\rm{dex}$. This enables us to mimic the star-forming galaxy selection applied in \cite{Best2021}, where radio-excess sources with luminosities exceeding the `ridgeline' value by $>0.7\,\rm{dex}$ are classified as radio-loud AGN and thus excluded from the SFG sample. Finally, we convert the modelled radio luminosities back to star formation rates, assuming Equation \ref{eq:Dan_relation}. This yields a sample of sources with estimates of star formation rates that are perturbed from their original values according to the modelled scatter on the $\rm{L_{150\,MHz}}-\rm{SFR}$ relation. \\
\indent We construct SFR functions for each instance of modelled scatter. We plot these in colour in Figure \ref{fig:L-SFR_cal}, also showing the true input SFR function in black. The modelled SFR functions differ substantially above $\sim\rm{SFR}_{\star}$: larger scatter in the relation causes the observed number density of galaxies to increase at the bright end, causing a gentler fall-off of the exponential (e.g. see the exaggerated scatter values of $0.5$ shown by the green lines). Because of the steepness of the original modelled SFR function, the number of sources `scattered up' to those SFRs (coloured lines) can vastly exceed (by up to $\sim1\,\rm{dex}$ at the highest SFRs) the genuine number of sources. This will have a significant impact on the SFR function derived from $\rm{L_{150\,MHz}}$ measurements, which we need to correct for. The differences we see are qualitatively in line with the differences seen in our observational estimates of SFR functions in Figure \ref{fig:L-SFR_cal_diffs}: as described in Section \ref{sec:sfrd_results}, the SFR functions constructed using consensus estimates tend to lie below those inferred directly from $L_{150\,\rm{MHz}}$ at $\rm{SFR}\gtrsim \rm{SFR}_{\star}$. \\
\indent In the right-hand panel of Figure \ref{fig:L-SFR_cal}, we plot the offsets between the original modelled SFR function and those modelled using various values of scatter in the $L_{150\,\rm{MHz}}-\rm{SFR}$ relation (i.e. deviation of each of the coloured lines from the black line). We use these modelled offsets, in combination with the differences between our multiple estimates of the SFR function, to constrain the true scatter on the $L_{150\,\rm{MHz}}-\rm{SFR}$ relation. We use the SFR functions derived from the consensus estimates as the `truth' (analogous to the black SFR function in the left-hand panel of Figure \ref{fig:L-SFR_cal}); of course, SFRs derived from SED fitting have their own uncertainties, but these are not dependent on the scatter in the $L_{150\,\rm{MHz}}-\rm{SFR}$ relation. We compare them to the SFRs derived from the radio luminosity functions, using a single $L_{150\,\rm{MHz}}-\rm{SFR}$ scaling with no scatter. Our data are shown in the right-hand panel of Figure \ref{fig:L-SFR_cal} (grey points). They are broadly consistent with a scatter of $\sim0.3\,\rm{dex}$ on the $L_{150\,\rm{MHz}}-\rm{SFR}$. This is in good agreement with the results of \cite{Smith2021}, who found significant scatter only at $\rm{SFR}>0$. They estimated the scatter to be $0.31 \pm 0.01\,\rm{dex}$ at $1 < \log_{10}(\rm{SFR}/\rm{M_{\odot}}\,\rm{yr^{-1}}) < 2$.\\
\indent As shown, the measured radio luminosity function (and the SFR function derived by scaling this with a single $L_{150\,\rm{MHz}}-\rm{SFR}$ calibration) will have an artificially shallower bright end slope due to the $\sim0.3\,\rm{dex}$ scatter in the $L_{150\,\rm{MHz}}-\rm{SFR}$ relation. This will lead to an overestimation of the cosmic star formation rate density. By integrating the modelled SFR function derived using different values of scatter, and comparing to the integrated base SFR function, we estimate the degree of boosting of the SFRD. We hence derive correction factors that can be applied to values of SFRD derived using a single $L_{150\,\rm{MHz}}-\rm{SFR}$ calibration. We derive the multiplicative correction factor $\rm{CORR_{SFRD}} = [1.0, 0.96, 0.93]$ for scatter $\Sigma=[0.2,0.3,0.4]\,\rm{dex}$. Without the exclusion of sources classified as radio-loud AGN, this correction would need to be larger. In Section \ref{sec:sfrd_results}, we correct our estimates of cosmic star formation rate density using a scaling factor of $0.96\pm0.04$, assuming a $\sim0.3\,\rm{dex}$ scatter in the $L_{150\,\rm{MHz}}-\rm{SFR}$ relation.

\begin{figure}
\includegraphics[width=\columnwidth]{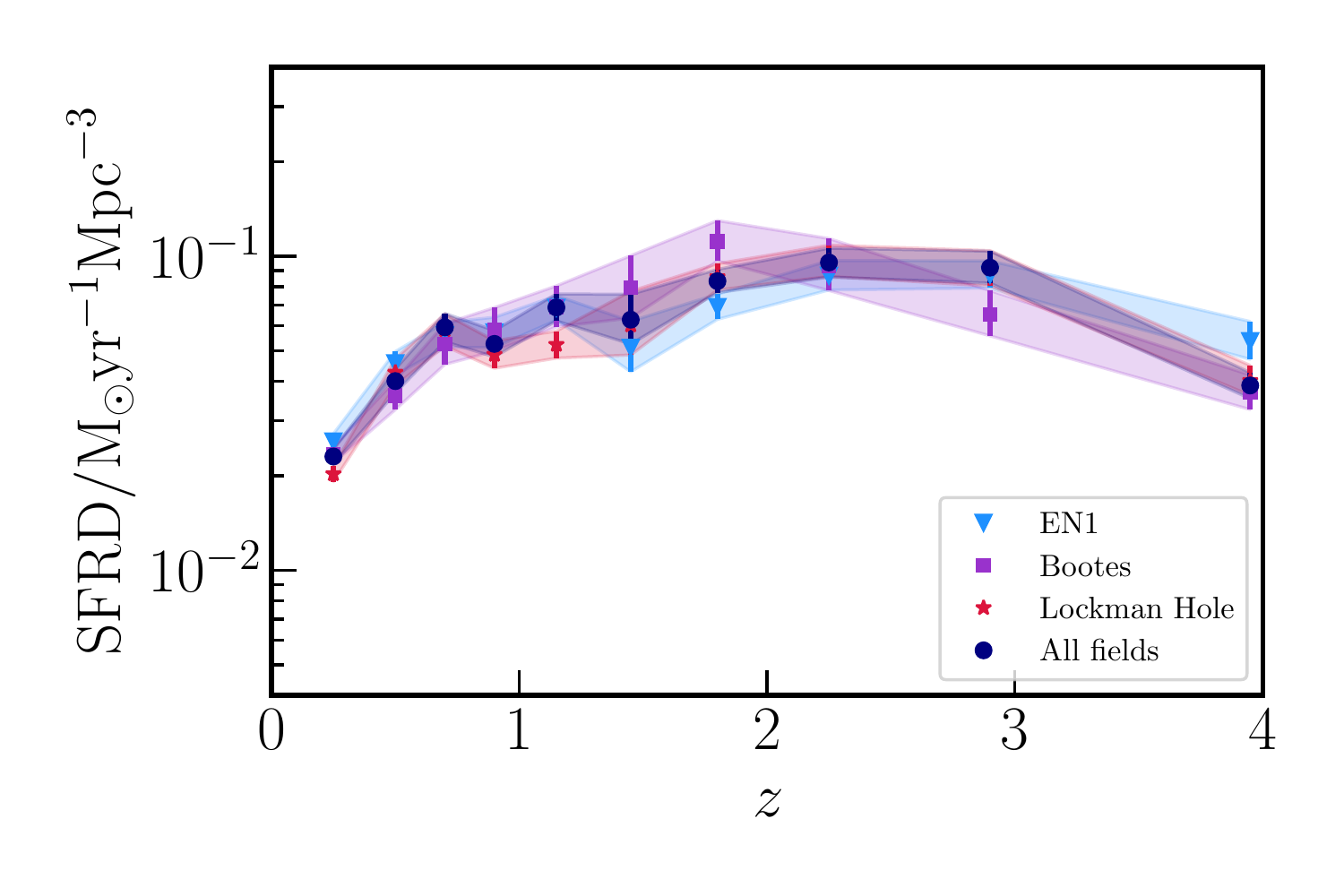}
\caption[]{The cosmic star formation rate density for each of the three LOFAR Deep Fields. These show good consistency, with larger uncertainties at $z\sim1.5$, where the lack of $H$-band data in Elais-N1 and the Lockman Hole drives particularly large uncertainties in photometric redshifts and at $z\gtrsim 3$, where our luminosity functions do not probe below $L_{\star}$. The data are tabulated in Table \ref{Table:sfrd_each_field}.}
\label{fig:sfrd_comparison}
\end{figure}

\begin{figure*}
\includegraphics[width=\columnwidth]{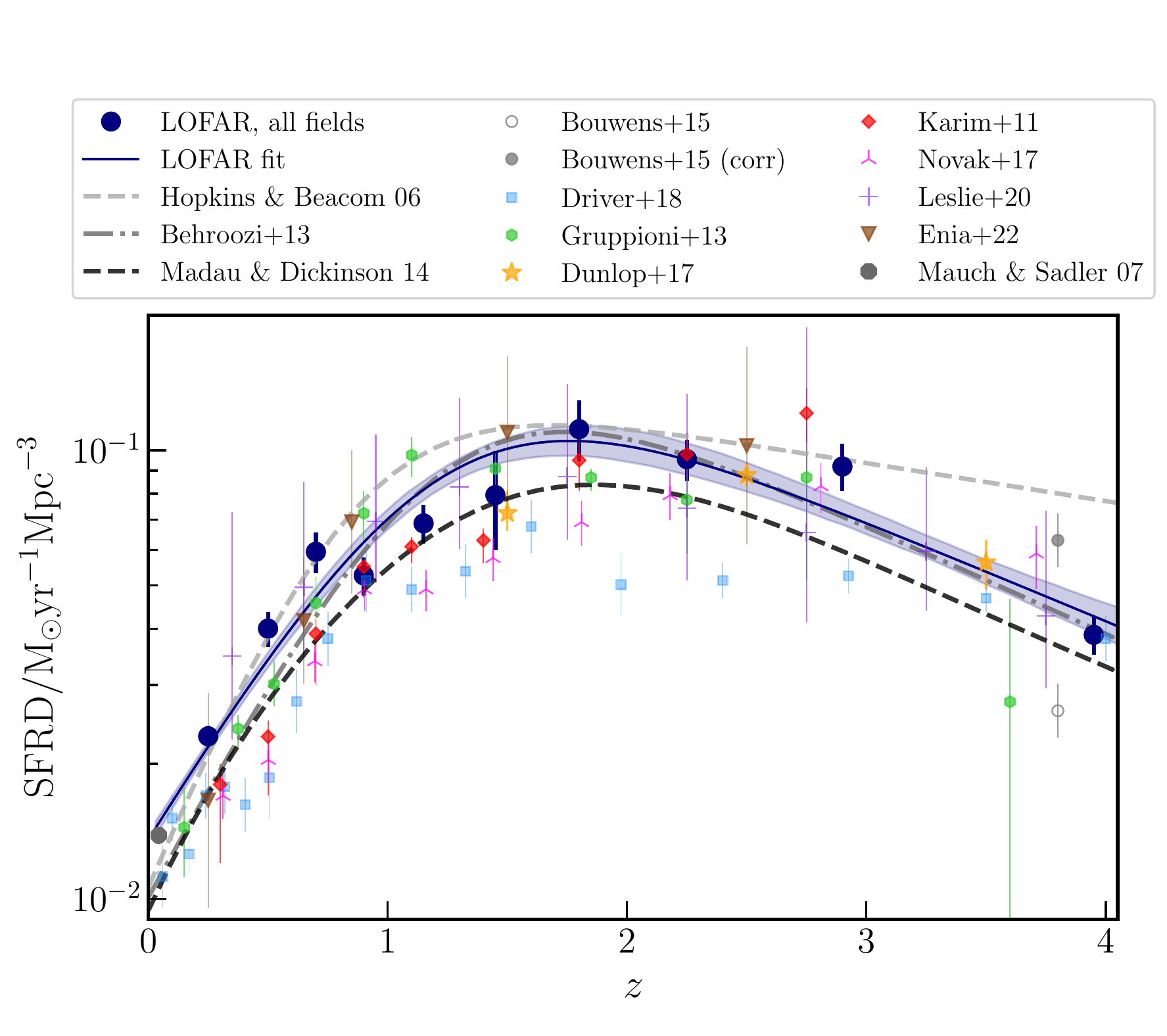}
\vspace{-0.8cm}
\caption[]{The cosmic star formation rate density, derived using the full sample of star-forming galaxies from LOFAR, with, literature data overlaid. Our new results using the three LOFAR deep fields combined are shown in navy (solid circles). At $z\sim1.5$, our best estimate of the SFRD comes from measurements taken in the Bo{\"o}tes field, due to increased uncertainties on the photometric redshifts of sources in Elais-N1 and the Lockman Hole. The navy shaded region shows our best estimate for the evolution of the SFRD from $z\sim0.2$ to $z\sim4$, the $1\,\sigma$ posterior of our fit to the SFRD derived from all three fields. We also include a $z\sim0$ measurement from \cite{Mauch2007}, shown in dark grey, in the fit. Our data are bracketed by previously-derived fits to data by \cite{Hopkins2006} (above) and \cite{Madau2014} (below); the \cite{Behroozi2013} fit shows best agreement. Coloured symbols represent a selection of estimates from the literature, derived using widely varying sample selections. These include UV-selected (\citealt{Bouwens2015a}; grey squares), optical/NIR-selected (blue squares; \citealt{Driver2018}), FIR-selected (green hexagons and orange stars; \citealt{Gruppioni2013} and \citealt{Dunlop2017}, respectively), and radio continuum-selected (red, pink, purple and brown symbols); \citealt{Karim2011,Novak2017,Leslie2020,Enia2022}). Overall, there is considerable scatter in measurements across the literature, with disagreements of $>0.4\,\rm{dex}$ at any given redshift. Our data show good agreement with the FIR-based measurements from \cite{Dunlop2017} and radio continuum-based analyses of \cite{Leslie2020} and \cite{Enia2022}. Measurements made by \cite{Driver2018} fall below our estimates at $z\lesssim3.5$.}
\label{fig:sfrd}
\end{figure*}

\section{The cosmic star formation density history}\label{sec:sfrd_results}
The cosmic star formation rate density can be calculated at a given epoch, z, by integrating the radio luminosity function as follows:
\begin{equation}\label{eq:sfrd_from_LF}
{\rm{SFRD}}(z) = {\rm{CORR_{SFRD}}}\times \int^{L_{\rm{max}}}_{L_{\rm{min}}}\phi(L,z)\,{\rm{SFR}}(L)\,{\rm{d}}\log_{10}L
\end{equation}
At each redshift, we derive estimates of the $\rm{SFRD}$ for each of the three fields and for all fields combined, by integrating the parameterised radio luminosity function presented in Equation \ref{eq:lf_fit} with the appropriate best-fitting parameters. We adopt a lower luminosity limit $L_{\rm{min}}=0.03\,L_{\star}$, and an upper luminosity limit of $L_{\rm{max}}=10^{28}\,\rm{W\,Hz^{-1}}$ (an order of magnitude brighter than our brightest luminosity bin). In Appendix \ref{sec:sfrd_integration_test} we present tests showing that the derived SFRD is robust to our choice of integration limits. $\rm{CORR_{SFRD}}=0.96\pm0.04$ is the correction factor derived in Section \ref{sec:L_sfr_cal}, which accounts for the potential bias in $\rm{SFRD}$ due to the scatter in the $L_{150\,\rm{MHz}}-\rm{SFR}$ relation. \\
\indent We present our estimates for the star formation rate density at $0\lesssim z\lesssim4$ in Figure \ref{fig:sfrd_comparison}. Estimates using the three fields individually show good consistency, with larger differences between fields at $z\sim1.5$, where the lack of $H$-band data in Elais-N1 and the Lockman Hole drives particularly large uncertainties in photometric redshifts (see Section \ref{sec:photo_z_uncertainty} and Figure \ref{fig:photoz_corrections}). Although we have attempted to resolve this by correcting for these uncertainties, the `corrected' data points for the SFRD at $z\sim1.5$ for Elais-N1 and the Lockman Hole disagree with the estimate using Bo\"{o}tes and with the neighbouring redshift bins (see Figure \ref{fig:sfrd_comparison} and Table \ref{Table:sfrd_each_field}). Because of this, we believe that these data are unreliable and we adopt Bo\"{o}tes data as our best estimate at $z\sim1.5$. There is also $\sim1\,\sigma$ discrepancy between the estimates derived from different fields at $z\sim3-4$. This arises due to the slightly different depths of the radio data. LOFAR coverage of Bo\"{o}tes is the shallowest, and fitting without the faintest luminosity data point leads to a slightly higher $L_{\star}$ and lower $\phi_{\star}$ being favoured.\\
\indent In Figure \ref{fig:sfrd} we plot our derived star formation rate density history from all three fields combined (see blue circles). We tabulate these estimates in Table \ref{Table:lf_with_z}. At most of the redshifts studied, our best estimate comes from integrating the luminosity function of the three fields combined: this gives the greatest numbers of sources, and enables us to average over any potential bias due to cosmic variance. At $z\sim1.5$, our best-estimate SFRD comes from measurements taken in the Bo{\"o}tes field, due to increased uncertainties on the photometric redshifts of sources in Elais-N1 and the Lockman Hole. 

\subsection{The functional form of the SFRD}
Following \cite{Hopkins2006} and \cite{Behroozi2013}, we use the {\it{emcee}} fitting code \cite{Foreman-Mackey2013} to fit the following functional form to our data, using the SFRD derived from all three LOFAR fields:
\begin{equation}
    {\rm{SFRD}}(z) = \frac{C}{10^{A(z-z_{0})}+10^{B(z-z_{0}})}.
\end{equation}
We also include an additional radio-derived SFRD measurement at $z=0.043$ \citep{Mauch2007} to anchor the fit at $z\sim0$ (note that we apply a correction to account for differing assumed IMFs). We adopt this measurement because the LoTSS Deep Fields do not probe enough volume to provide secure cosmic SFRD measurements at $z\sim0$; in comparison, the \citep{Mauch2007} measurement is made over a $\sim300$ times larger sky area ($7000\,\rm{deg}^2$). Although too shallow for high-redshift studies like ours, their data are deep enough to constrain the SFRD at very low redshift.
We derive $A=-0.89^{+0.08}_{-0.07}$, $B=0.22\pm0.04$, $\log_{10}C=-0.76\pm0.05$ and $z_{0}=1.22\pm0.15$. \\
\indent The best-fitting SFRD is overplotted and compared to previous fits in Figure \ref{fig:sfrd}. Our data and fit are broadly consistent with previous fits to older data, lying approximately at or below the estimate of \cite{Hopkins2006} but above that of \cite{Madau2014} by $\lesssim 0.1\,\rm{dex}$ at $0 \lesssim z \lesssim 4.0$. Notably, \cite{Leslie2020} also find that their radio-derived SFRD estimates lie $\lesssim 0.15\,\rm{dex}$ above \cite{Madau2014} at $z\lesssim 3$. We are in excellent agreement with the fit presented by \cite{Behroozi2013}; their fitted form lies within $\sim1\,\sigma$ of our fit at $0.7\lesssim z \lesssim 4.0$. At lower redshifts, we measure a shallower evolution of the SFRD.

\subsection{Comparison to literature data}
In this section, we compare our SFRD measurements to estimates from individual studies of star-forming galaxies selected using a variety of methods. While there exist a vast number of low-z measurements in the literature, we focus on those that reach out to high redshift, including several that have been published since the compilation of \cite{Madau2014}, and those that use longer-wavelength SFR estimators. As shown in Figure \ref{fig:sfrd}, previous estimates show a consistent general form, with the SFRD increasing between $z=0$ and $z\sim2$ and then declining towards higher redshift. However, exact measurements disagree by $>0.4\,\rm{dex}$ at any given redshift. Here, we compare our LOFAR results to several previous measurements in detail. \\
\indent \cite{Bouwens2015a} identified galaxies at $z\sim4-10$ in the {\it HST} legacy fields using the Lyman break technique. At $z=3.8$, their sample consists of $B-$band dropouts. They estimated the SFRD from the raw UV luminosities and also using a correction for dust attenuation based on the IRX-$\beta$ relation. At $z=3.8$, the mean dust extinction, $A_{\rm{UV}}$, is $2.4$. We correct their estimates from a \cite{Salpeter1955} to \cite{Chabrier2003} IMF, and plot both corrected and uncorrected estimates in grey. Our dust-independent estimate lies in between the dust-uncorrected and dust-corrected values, which may indicate that the necessary correction for dust attenuation was overestimated. \\
\indent \cite{Driver2018} combined $r$-band selected galaxies from GAMA \citep{Driver2011,Liske2015}, $i$-band selected galaxies from G10-COSMOS \citep{Davies2015,Andrews2017} and $1.6\,\mu\rm{m}$-selected galaxies from 3D-{\it HST} \citep{Momcheva2016} in their analysis. They derived SFRs using {\small{MAGPHYS}}, with various combinations of multi-wavelength data. Their SFRD estimates (blue squares) fall below ours at all redshifts apart from $z\sim3.5$. Interestingly, there is a particular increase in the offset between $z=1.6$ (offset $0.18\,\rm{dex}$) and $z=1.975$ (offset $0.31\,\rm{dex}$). Between these two redshifts, the sample changes from including G10-COSMOS sources to a 3D-{\it HST}-only sample for which SFRs are derived without FIR data. The large offsets between our dust-independent estimate and theirs at $z\sim2$ and $z\sim2.4$ suggests that dust-obscured star formation is significant at these redshifts and that, in the absence of FIR data, they are under-predicting the SFRD. Our estimates are in near-perfect agreement at $z\sim3.5-4$. This implies that the contribution of dust-obscured star formation to the total SFRD is lower by then; this is broadly in agreement with \cite{Dunlop2017} and \cite{Zavala2021}.\\
\indent SFRD estimates using FIR-based studies are generally in better agreement with ours. We overplot estimates from {\it Herschel}-selected samples (\citealt{Gruppioni2013}; green hexagons). These agree with our best-fitting line to within $\sim0.15\,\rm{dex}$ at $z<3.5$. At $z\sim3.6$, their error bars are very large but remain consistent with our estimate. We also compare to measurements from \cite{Dunlop2017} (orange stars), who combined direct detections and stacking of ALMA imaging in the {\it Hubble Ultra Deep Field} with {\it{HST}}-derived SFR measurements from the rest-frame UV to estimate the total SFRD. Our measurements are consistent with theirs. \\
\indent Finally, we compare to other work based on radio continuum emission. \cite{Karim2011} stacked $1.4\,\rm{GHz}$ data from VLA-COSMOS, at positions of a $3.6\,\mu\rm{m}$-selected sample of $>10^{5}$ galaxies. \cite{Leslie2020} built on this work, stacking $3\,\rm{GHz}$ data within the same field and refining the source selection to a fully mass-selected sample (using $K_{S}$-band data for galaxies at $z<2.5$ and $3.6\,\mu\rm{m}$ data at higher redshifts). Our measurements are fully consistent with those of \cite{Leslie2020}, but are discrepant with those of \cite{Karim2011} at $z\lesssim1.5$. \cite{Leslie2020} note that \cite{Karim2011} use a different $L_{\rm{radio}}-$SFR calibration, which yields lower SFRs, but differences are also likely driven by the deeper parent catalogue used by \cite{Leslie2020}. \\
\indent \cite{Enia2022} constructed a $1.4\,\rm{GHz}$-selected sample using VLA observations in GOODS-N to measure the SFRD to $z\sim3.5$. These measurements are also fully consistent with ours and with those of \cite{Leslie2020}. We note, though, that our considerably larger area ($\sim26\,\rm{deg}^{2}$ compared to $\sim2\,\rm{deg}^{2}$ for VLA-COSMOS and $171\,\rm{arcmin}^{2}$ for GOODS-N) enables much tighter constraints. The measurements of \cite{Novak2017} are below ours (by up to $0.4\,\rm{dex}$) and also below those of \cite{Enia2022} and \cite{Leslie2020}. This is perhaps surprising, given that like \cite{Enia2022} and this work, \cite{Novak2017} use a radio-selected sample and the luminosity functions they derive are in good agreement with ours (once scaled to the same rest-frame wavelength; see Figure \ref{fig:all_rq_lfs}). \cite{Enia2022} suggest that discrepancies between their measurements and those of \cite{Novak2017} might be due to the shallower faint end slope used by \cite{Novak2017}, but our fitted faint end slope is actually slightly shallower than that derived by \cite{Novak2017} (note that \citealt{Novak2017} derive their slope from other samples of radio data rather than their own VLA-COSMOS data; at the faint end, data in their fit is drawn from \citealt{Condon2002}). As noted by \cite{Leslie2020} and highlighted in Figure \ref{fig:L-SFR_cal_diffs}, the impact of different $L_{\rm{radio}}-$SFR calibrations is significant. This can be seen most clearly in the differences between our SFRD predictions and those derived by \cite{Novak2017}. Given the consistency with our luminosity function measurements out to $z\sim3$, this discrepancy appears to stem from their different, redshift-dependent, $q_{\rm{IR}}$-based $L_{\rm{radio}}-$SFR conversion.\\
\indent Our results highlight significant differences between SFRD measurements derived from UV/optical/IR data and those derived from FIR/radio data. As described above, these differences likely stem from a number of sources. Incomplete samples and uncertainties in dust corrections affect samples selected at shorter wavelengths, and may drive some of the differences between the SFRD estimated by \cite{Driver2018} and other estimates. Differences in the adopted SFR calibrations and values assumed for the faint-end slope of luminosity functions affect all estimates, and are most clearly seen from the different SFRD measurements derived using similar data (e.g. discrepancies between the $1.4\,\rm{GHz}$-derived SFRD measurements constructed by \citealt{Karim2011} \citealt{Novak2017}, \citealt{Leslie2020}, and \citealt{Enia2022}). In this work, we fix the faint-end slope of the radio luminosity function to the value derived at $z=0.03-0.30$. Its true value is unconstrained by our data at higher redshifts, and evolution would lead to systematic errors in our SFRD estimates. For example, \cite{Yuksel2008} noted that a steeper faint-end slope at high redshift could help reconcile SFRD estimates made by integrating UV luminosity functions with those made using gamma ray bursts.

\begin{figure}
\includegraphics[width=\columnwidth]{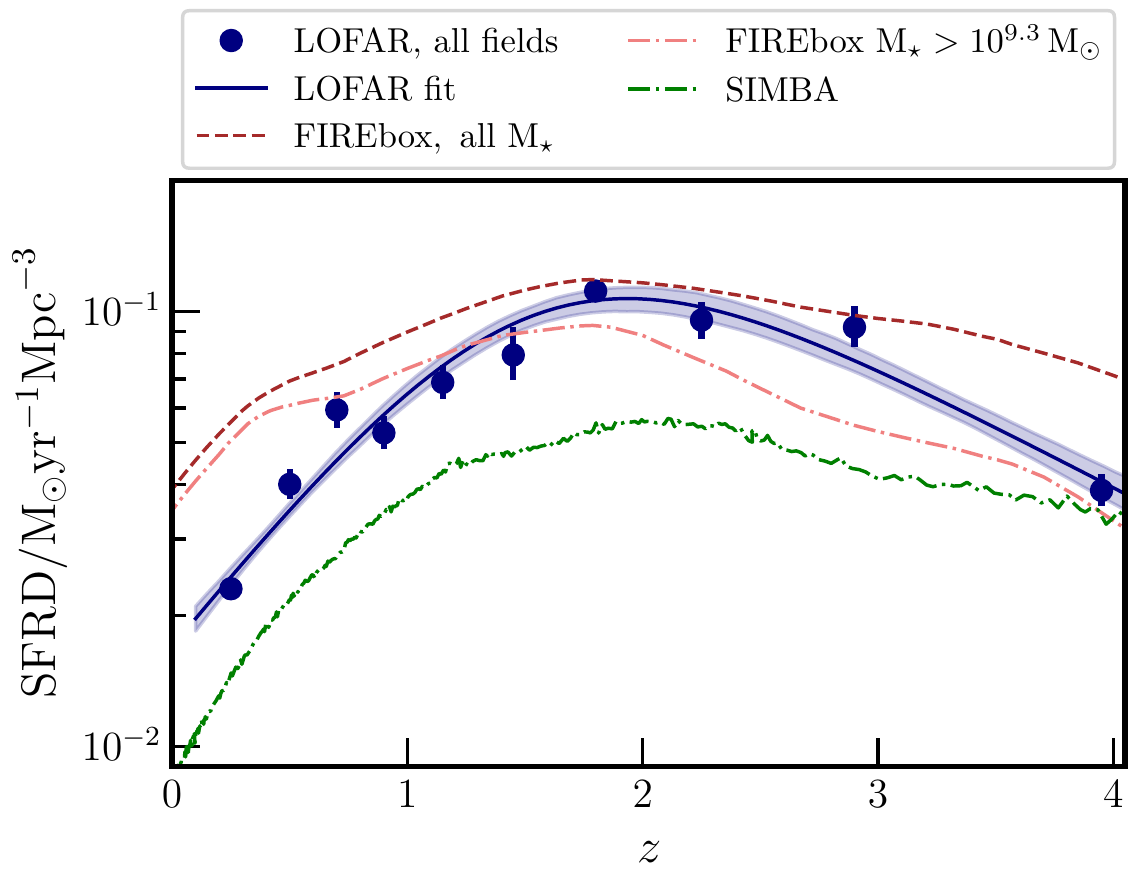}
\caption[]{The LOFAR-derived cosmic star formation rate density presented in Figure \ref{fig:sfrd} (navy stars, with fitted functional form in shaded navy), with predictions from hydrodynamic simulations overlaid. The total SFRD predicted by FIREBox (dark red; \citealt{Feldmann2022}) exceeds our estimate, with the greatest deviations at $z<1$ and $z>3$. When the FIREbox SFRD is calculated using only galaxies with $M_{\star}>10^{9.3}\rm{M_{\odot}}$ (pale red; this limit corresponds approximately to $0.03\,L_{\star}$, which we integrate down to in this work), agreement is better at $z\sim1-2$. At lower redshifts, the FIREbox SFRD estimate is up to a factor of a few higher than our data, likely due to the lack of AGN feedback in the simulations. S{\sc imba} \citep{Dave2019} under-predicts the SFRD at all redshifts, with the most substantial discrepancies at $z\sim2$.}
\label{fig:sfrd_models}
\end{figure}

\subsection{Comparison to models of galaxy formation}
We show predictions for the SFRD from various models of galaxy formation in Figure \ref{fig:sfrd_models}. We note here that a proper comparison would involve making predictions for the multi-wavelength emission (including radio continuum) of the simulated sources, folding in source detection and classification based on the mock SEDs and repeating the analysis on analogously-selected galaxy samples. This is clearly beyond the scope of this paper; instead we present a brief comparison with some initial thoughts here.\\
\indent FIREbox \citep{Feldmann2022} evolves a small cosmological volume $(22.1\,\rm{cMpc})^{3}$ down to $z=0$ using the models initially designed for zoom-in galaxies within the `Feedback in Realistic Environments' (FIRE) project \citep{Hopkins2014,Hopkins2017,Hopkins2022}. FIREbox represents a simulation with particularly high dynamic range, given its low baryonic particle mass ($M_{\rm{baryon}}=6\times10^{4}\,\rm{M_{\odot}}$) and medium box size. The total volume-averaged star formation rate density lies above our estimate at all redshifts, with particular deviations at $z<1$ and $z>3$. This could be due to a number of factors, including galaxy selection effects. When a stellar mass selection of $M_{\star}>10^{9.3}\,\rm{M_{\odot}}$ (approximately corresponding to $L>0.03\,L_{\star}$) is applied to the simulated galaxies, the predicted SFRD changes significantly, showing better agreement with our data at $z\sim1-2$ and a more rapid decrease at $z\gtrsim2$. Importantly, the difference between FIREbox predictions and our data at high-$z$ is largely associated with low mass, lower SFR objects that fall below the detection limit of current radio surveys. Deeper radio surveys, or potentially radio-stacking approaches \citep[e.g.][]{Leslie2020}, may bring the data into better agreement with the simulation. At low redshifts, FIREbox underestimates the fraction of massive, quenched galaxies (\citealt{Feldmann2022}, see also \citealt{Parsotan2021}); this is likely responsible for the overestimation of the SFRD relative to our estimate at $z\lesssim 1$. The inclusion of AGN feedback in the simulation may help suppress star formation and alleviate this (Cochrane et al. in preparation).\\
\indent We also compare our estimates to data from the cosmological hydrodynamical simulation S{\sc imba} \citep{Dave2019}. SFRD estimates from S{\sc imba} include all the star formation in the box, with no cut on galaxy stellar mass. As noted by \cite{Dave2019}, the S{\sc imba}-predicted SFRD peaks slightly earlier than the compilation of \cite{Madau2014}; since our LOFAR estimate of the position of the peak is similar to that of \cite{Madau2014}, the S{\sc imba} SFRD also peaks earlier than our LOFAR estimate. At all redshifts, S{\sc imba} predicts a lower SFRD than measured from LOFAR, with the most substantial discrepancies (of more than a factor of $2$) seen at $z\sim2$. This is consistent with the star-forming main sequence of S{\sc imba} galaxies displaying a lower normalisation than is observed at this epoch.

\subsection{Further work}
In this work, we have provided important constraints on the cosmic history of star formation based on a statistical study of the deepest available low-frequency radio source counts. However, there are some areas in which our method might be improved in future work. We have adopted a simplified model of the separation of radio sources into those produced by AGN and those produced by stars; in reality, there will be sources where both mechanisms contribute to the radio emission. There will be a jet contribution to the radio emission in some of the radio-quiet AGN included in the sample in this paper, and a star formation contribution to the radio emission in some of the radio-loud AGN excluded from this work. It has long been known that synchrotron radio jets ejected by the AGN can induce star formation as they propagate outwards from their host galaxy nuclei into the galactic and intergalactic medium \citep[e.g.][]{Rees1989,Gaibler2012}. A prominent low-redshift example is 3C277.3/Coma A \citep{Miley1981,Capetti2022}. There are also indications that this mechanism could be important at high redshift; in both the spiderweb proto-cluster at $z = 2.2$ and 4C41.17 at $z = 3.8$, alignments seen between the radio, optical, CO and X-ray emission have been interpreted as star formation being induced by the radio jets (\citealt{Bicknell2000,Miley2008}; see also \citealt{Duncan2023}). \\
\indent LOFAR's unique combination of sensitivity and high resolution at low frequencies equips it well to detect and map radio-loud galaxies out to the highest redshifts. Imaging with the international baselines will help to distinguish radio jets from star formation morphologically \citep[e.g.][]{Morabito2022}. In addition, the new William Herschel Telescope Enhanced Area Velocity Explorer (WEAVE; \citealt{Jin2022a}), a multi-object fiber-fed spectrograph that has just seen first light, will target all radio-detected sources within the LOFAR Deep Fields \citep[WEAVE-LOFAR;][]{Smith2016}. This will provide a vastly larger number of spectroscopic redshifts for the radio sources, additionally enabling better source classifications.

\section{Conclusions}\label{sec:conclusions}
In this paper, we have used data from the pioneering wide and deep LOFAR Deep Fields to study the cosmic star formation history in a dust-independent manner. The three fields studied, Elais-N1, Bo{\"o}tes and the Lockman Hole, all benefit from extensive UV-FIR coverage, enabling the reliable exclusion of AGN-dominated radio sources from our analysis. We derive $150\,\rm{MHz}$ luminosity functions for samples of galaxies with radio emission dominated by star formation, from $z\sim0$ to $z\sim4$. Our main conclusions are summarised here:
\begin{itemize}
\item Out to $z\sim3$, our $150\,\rm{MHz}$ luminosity functions are in good agreement with the scaled $1.4\,\rm{GHz}$ luminosity functions derived by \cite{Novak2017} using VLA-COSMOS data (assuming a spectral index $\alpha=-0.7$). Given the larger area spanned by the LOFAR Deep Fields ($\sim25\,\rm{deg}^{2}$, compared to the $\sim2\,\rm{deg}^{2}$ VLA-COSMOS survey), and the use of three fields to overcome cosmic variance, we can constrain radio luminosity functions to roughly an order of magnitude brighter luminosities, while reaching similar luminosities at the faint end. 
\item Our derived $0<z<0.3$ $150\,\rm{MHz}$ luminosity function is well-fitted by a parametrisation of the form: \\ $\phi(L) = \phi_{\star}\Bigg(\frac{L}{L_{\star}}\Bigg)^{1-\alpha}\exp\Bigg[-\frac{1}{2\sigma^{2}}\log^{2}\Bigg(1+\frac{L}{L_{\star}}\Bigg)\Bigg]
$, \\
with $\log_{10}(\phi_{\star}/\,\rm{Mpc^{-3}\,dex^{-1}})=-2.46\pm0.01$, $\sigma=0.49\pm0.01$, $\alpha=1.12\pm0.01$ and $\log_{10}(L_{\star}/\mathrm{W\,Hz}^{-1})=22.40^{+0.02}_{-0.03}$.
\item Using the values of $\sigma$ and $\alpha$ derived to our low redshift data, we fit our higher redshift radio luminosity functions using the same parametrisation, to constrain the evolution of $\phi_{\star}$ and $L_{\star}$.
\item We show that transforming a radio luminosity function to a star formation rate function is complicated by the scatter in the $L_{\rm{radio}}-\rm{SFR}$ relation. Star formation rate functions derived using this  conversion tend to lie above those derived using SFRs obtained from SED fitting at $\rm{SFR}\gtrsim\rm{SFR_{\star}}$, with deviations of up to an order of magnitude at the highest SFRs. Using a simple model, we show that higher values of scatter in the $L_{\rm{radio}}-\rm{SFR}$ cause a gentler fall-off of the exponential of the measured radio luminosity function. This effect is most important where the luminosity (or star-formation rate) function is steepest, above its break. This can lead to an overestimation of the cosmic star formation rate density, which is generally derived by integrating the measured luminosity function. The magnitude of the correction depends on the the form of the luminosity function, the scatter in the $L_{\rm{radio}}-\rm{SFR}$ relation, and the details of the sample selection (i.e. whether sources that are particularly radio-loud for their SFR are excluded from the sample). By comparing the difference in the inferred SFRFs using the two methods to our model of the bias, we constrain the scatter in the $L_{\rm{radio}}-\rm{SFR}$ relation of star-forming galaxies to be $\sim0.3\,\rm{dex}$. Encouragingly, this value is in line with recent work that constrains the scatter using an independent method \citep{Smith2021}. We derive an appropriate correction factor to apply to the SFRD of $\sim0.96\pm0.04$.
\item We constrain the cosmic star formation rate density from $z\sim0$ to $z\sim4$, by integrating our $L_{150\,\rm{MHz}}$ luminosity functions, in combination with a self-consistently derived $L_{150\,\rm{MHz}}-\rm{SFR}$ relation, correcting for its scatter. Since the SFRD is constructed using radio-selected samples, our measurements are robust to the effects of dust. Our derived SFRD lies between previous compilations at all redshifts studied. Our measurements are in good agreement with those previously derived using smaller $1.4\,\rm{GHz}$-selected samples \citep[e.g.][]{Leslie2020,Enia2022} and from FIR-based studies \citep[e.g.][]{Gruppioni2013,Dunlop2017}. Our derived SFRD is well-fitted by a model of the form ${\rm{SFRD}}(z) = \frac{C}{10^{A(z-z_{0})}+10^{B(z-z_{0}})}$, with $A=-0.89^{+0.08}_{-0.07}$, $B=0.22\pm0.04$, $\log_{10}C=-0.76\pm0.05$ and $z_{0}=1.22\pm0.15$.
\end{itemize}

\indent Prospects for future census studies of radio-selected star-forming galaxies are bright. The LOFAR Deep Fields survey continues to observe all three fields. EN1 has already been observed for $500\,\rm{hr}$, with imaging reaching $\sim12\,\mu\rm{Jy/beam}$. By mid-2023 (following LOFAR Cycles 18 and 19), we expect to reach $16\,\mu\rm{Jy/beam}$ in Bo{\"o}tes with $312\,\rm{hr}$ of data, and $13\,\mu\rm{Jy/beam}$ in the Lockman Hole with $352\,\rm{hr}$ of data. We are also observing in the NEP, where we expect to reach $13\,\mu\rm{Jy/beam}$ in $400\,\rm{hr}$. These deeper LoTSS radio data, alongside spectra from WEAVE-LOFAR, will build up large, dust-independent samples of star-forming galaxies for further study, at fainter star formation rates than previously possible. This will enable not only the characterisation of the global cosmic SFRD but also the investigation of the drivers of star formation and quenching in sub-populations over cosmic time.
\section*{Acknowledgements}
We thank the anonymous referee for very helpful comments. This paper is based (in part) on data obtained with the International LOFAR Telescope (ILT) under project codes LC0 015, LC2 024, LC2 038, LC3 008, LC4 008, LC4 034 and LT10 01. LOFAR \citep{VanHaarlem2013} is the Low Frequency Array designed and constructed by ASTRON. It has observing, data processing, and data storage facilities in several countries, which are owned by various parties (each with their own funding sources), and which are collectively operated by the ILT foundation under a joint scientific policy. The ILT resources have benefited from the following recent major funding sources: CNRS-INSU, Observatoire de Paris and Universit\'e d’Orl\'eans, France; BMBF, MIWF-NRW, MPG, Germany; Science Foundation Ireland (SFI), Department of Business, Enterprise and Innovation (DBEI), Ireland; NWO, The Netherlands; The Science and Technology Facilities Council, UK; Ministry of Science and Higher Education, Poland.\\
\indent RKC is grateful for the support of a Flatiron Research Fellowship, John Harvard Distinguished Science Fellowship and the Institute for Astronomy, University of Edinburgh. The Flatiron Institute is supported by the Simons Foundation. RK acknowledges support from the STFC through grant ST/V000594/1. PNB and JS are grateful for support from the UK STFC via grants ST/R000972/1 and ST/V000594/1. DJBS and MJH acknowledge support from the UK STFC via grant ST/V000624/1. IP acknowledges support from INAF under the SKA/CTA PRIN “FORECaST“ and PRIN MAIN STREAM “SAuROS” projects and under the Large Grants 2022 programme. KJD acknowledges funding from the European Union’s Horizon 2020 research and innovation programme under the Marie Sk\l{}odowska-Curie grant agreement No. 892117 (HIZRAD).\\
\indent RKC thanks Robert Feldmann and Romeel Dav\'{e} for sharing data from FIREBox and S{\sc imba}, respectively.

\section*{Data availability}
The data used in this paper are derived from the LoTSS Deep Fields Data Release 1 \citep{Tasse2021,Sabater2021,Duncan2021,Kondapally2021}. The images and catalogues are publicly available at \url{https://lofar-surveys.org/deepfields.html}. Other results presented in the paper are available upon reasonable request to the corresponding author.

\bibliographystyle{mnras}
\bibliography{Edinburgh}


\appendix
\section{Radio completeness corrections}
In Figure \ref{fig:completeness_sfg}, we show radio completeness as a function of $150\,\rm{MHz}$ flux density for each LOFAR Deep Field. We tabulate these values in Table \ref{Table:completeness}.
\begin{table}
\begin{center}
\begin{tabular}{cccc}
\hline
Flux density/mJy & Elais-N1 & Lockman Hole & Bo\"{o}tes \\
\hline
0.11 & 0.227 & - & - \\
0.13 & 0.371 & 0.184 & - \\
0.16 & 0.586 & 0.301 & - \\
0.19 & 0.741 & 0.423 & 0.261 \\
0.23 & 0.836 & 0.599 & 0.324 \\
0.28 & 0.889 & 0.742 & 0.5 \\
0.33 & 0.918 & 0.846 & 0.669 \\
0.4 & 0.939 & 0.891 & 0.806 \\
0.63 & 0.963 & 0.945 & 0.913 \\
1.01 & 0.97 & 0.955 & 0.954 \\
1.59 & 0.977 & 0.977 & 0.972 \\
2.52 & 0.985 & 0.981 & 0.98 \\
4.0 & 0.979 & 0.977 & 0.98 \\
6.34 & 0.984 & 0.973 & 0.986 \\
10.05 & 0.983 & 0.981 & 0.989 \\
15.92 & 0.984 & 0.985 & 0.989 \\
25.24 & 0.984 & 0.986 & 0.992 \\
40.0 & 0.988 & 0.984 & 0.991 \\
\end{tabular}
\caption{Radio completeness at flux densities in the range $0.11-40\,\rm{mJy}$, for each field studied. The source size distribution used was defined using the size distribution of star-forming galaxies with flux densities in the range $1-5\,\rm{mJy}$. A similar table for a source size distribution appropriate for AGN is presented by \protect\cite{Kondapally2021a}.}
\label{Table:completeness}
\end{center}
\end{table}

\section{Photometric uncertainty corrections}
In Section \ref{sec:photo_z_uncertainty}, we describe corrections derived to account for uncertainties in photometric redshifts. These corrections are shown for each field in Figure \ref{fig:photoz_corrections} and tabulated in Table \ref{Table:photoz_correction_table}.

\begin{table}
\begin{center}
\begin{tabular}{cccc}
\hline
Redshift & Elais-N1 & Lockman Hole & Bo\"{o}tes \\
\hline
$0.1-0.4$ & $0.97$ & $1.00$ & $0.98$ \\
$0.4-0.6$ & $0.91$ & $0.98$ & $0.99$ \\
$0.6-0.8$ & $1.02$ & $0.97$ & $1.00$ \\
$0.8-1.0$ & $0.91$ & $1.04$ & $1.00$  \\
$1.0-1.3$ & $1.18$ & $1.05$ & $0.97$  \\
$1.3-1.6$ & $0.65$ & $0.63$ & $0.88$ \\
$1.6-2.0$ & $1.34$ & $1.35$ & $1.22$  \\
$2.0-2.5$ & $1.16$ & $0.99$ & $1.05$  \\
$2.5-3.3$ & $1.14$ & $1.07$ & $1.09$  \\
$3.3-4.6$ & $1.00$ & $1.00$ & $1.00$  \\
\end{tabular}
\caption{Photometric correction factors, for each field, as described in Section \ref{sec:photo_z_uncertainty} and shown in Figure \ref{fig:photoz_corrections}.}
\label{Table:photoz_correction_table}
\end{center}
\end{table}

\section{SFRD estimates for each field}
Estimates for the star formation rate density for each individual LOFAR Deep Field are presented in Table \ref{Table:sfrd_each_field}.

\begin{table}
\begin{center}
\begin{tabular}{l|c|c|c}
\hline
Redshift & \multicolumn{3}{c}{$\rm{SFRD}/M_{\odot}\,\rm{yr}^{-1}\,\rm{Mpc}^{-3}$}\\
\hline
& Elais-N1 & Lockman Hole & Bo{\"o}tes \\ 
\hline
$0.1-0.4$ & $0.026\pm0.001$ & $0.020\pm0.001$ & $0.023\pm0.001$ \\
$0.4-0.6$ & $0.046\pm0.003$ & $0.043\pm0.004$ & $0.036\pm0.003$ \\
$0.6-0.8$ & $0.056\pm0.004$ & $0.058\pm0.006$ & $0.053\pm0.007$ \\
$0.8-1.0$ & $0.057\pm0.006$ & $0.049\pm0.004$ & $0.058\pm0.009$ \\
$1.0-1.3$ & $0.069\pm0.005$ & $0.052\pm0.005$ & $0.069\pm0.010$ \\
$1.3-1.6$ & $0.051\pm0.009$ & $0.060\pm0.014$ & $0.079\pm0.018$ \\
$1.6-2.0$ & $0.069\pm0.005$ & $0.086\pm0.007$ & $0.111\pm0.017$ \\
$2.0-2.5$ & $0.087\pm0.008$  & $0.096\pm0.010$ &$0.093\pm0.017$\\
$2.5-3.3$ & $0.087\pm0.008$ & $0.091\pm0.011$ & $0.065\pm0.010$ \\
$3.3-4.6$ & $0.054\pm0.007$ & $0.040\pm0.004$ & $0.037\pm0.004$ \\
\end{tabular}
\caption{Cosmic star formation rate density estimates for the three individual fields, as shown in Figure \ref{fig:sfrd_comparison}.}
\label{Table:sfrd_each_field}
\end{center}
\end{table}

\section{The impact of systematic corrections on the derived cosmic star formation rate density}
\begin{figure*}
\includegraphics[width=\columnwidth]{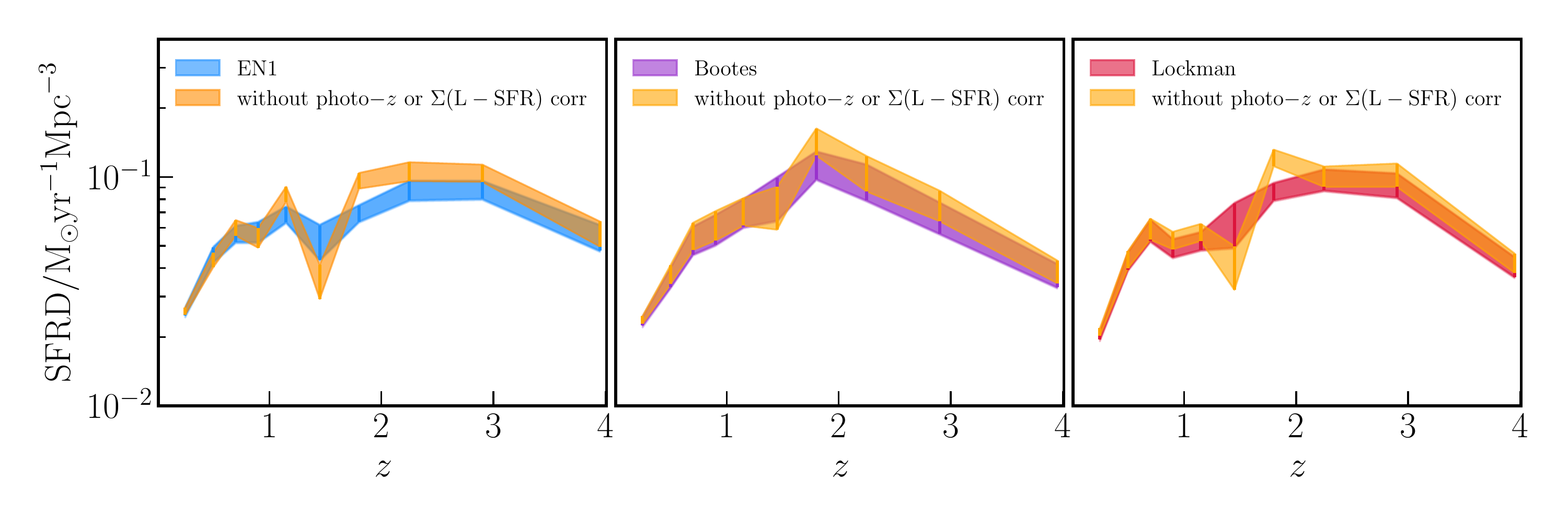}
\caption[]{The cosmic star formation rate density, derived by integrating the radio luminosity function using the $L_{150\,\rm{MHz}}-\rm{SFR}$ relation of \cite{Smith2021}, is shown in blue, purple and red for Elais-N1, Bo{\"o}tes and the Lockman Hole, respectively. In orange, we show the SFRD that would be derived without correcting for uncertainties in the photometric redshifts (see Figure \ref{fig:photoz_corrections} and Table \ref{Table:photoz_correction_table}) or scatter in the $L_{150\,\rm{MHz}}-\rm{SFR}$ relation (see Section \ref{sec:L_sfr_cal}). Corrections for photometric redshift uncertainties have minor effects on the majority of redshift bins, but are important at $z\sim1-2$ for Elais-N1 and the Lockman Hole, which lack $H-$band data.}
\label{fig:sfrd_3panel}
\end{figure*}

In Figure \ref{fig:sfrd_3panel}, we show the impact of applying corrections for uncertainties in photometric redshifts and the scatter in the $L_{150\,\rm{MHz}}-\rm{SFR}$ relation on the derived SFRD for each of the three fields. Corrections for photometric redshift uncertainties lead to small changes in the estimated SFRD for the majority of redshift bins, but are important at $z\sim1-2$ for Elais-N1 and the Lockman Hole, the two fields lacking $H$-band data. The correction for the scatter in the $L_{150\,\rm{MHz}}-\rm{SFR}$ relation serves to move the whole derived relation to slightly lower SFRD values.

\section{The impact of integration limits on the derived SFRD}\label{sec:sfrd_integration_test}
\begin{figure*}
\includegraphics[width=\columnwidth]{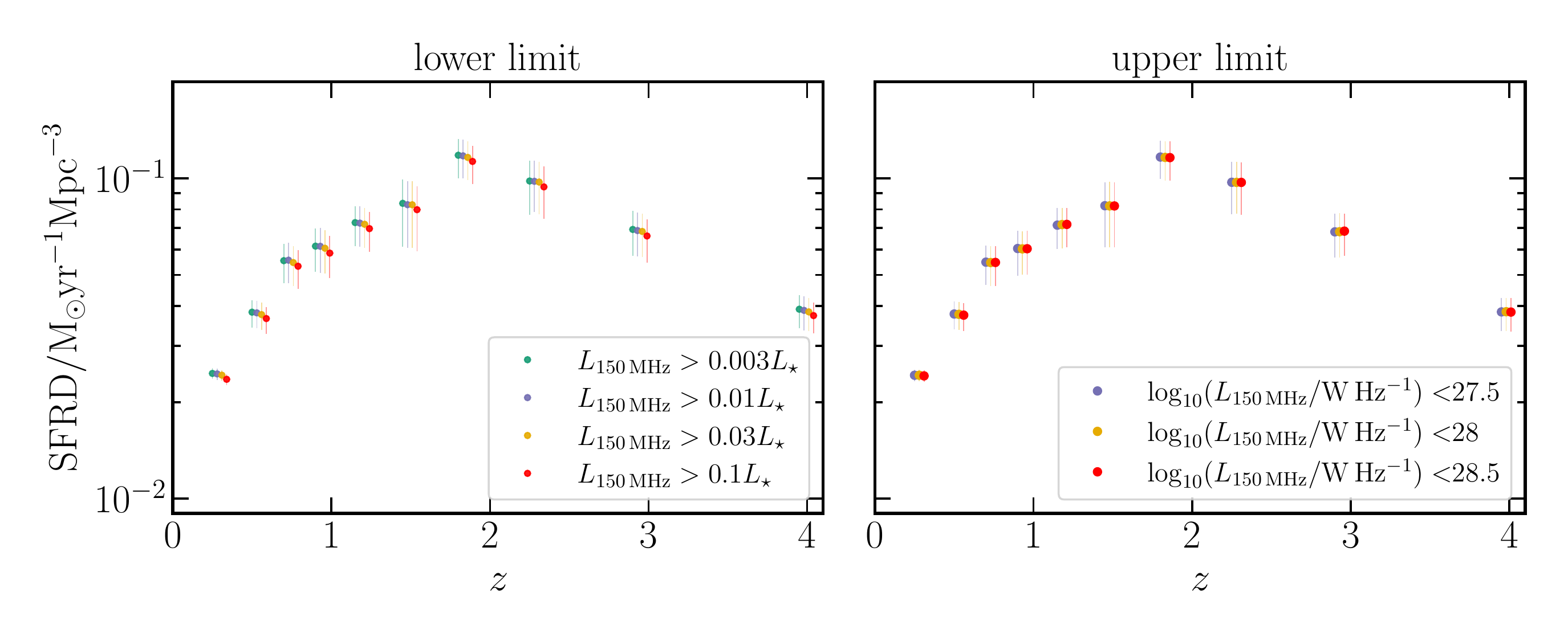}
\caption[]{The impact of the choice of lower (left) and upper (right) integration limit on the derived SFRD, shown here for Bo{\"o}tes. Small artificial x-axis offsets are added to display small differences most clearly. The impact of changing either the lower or the upper integration limit is small compared to the reported uncertainties.}
\label{fig:integration_test}
\end{figure*}

We have tested the impact of changing the range of radio luminosities over which we integrate to obtain the SFRD. In Figure \ref{fig:integration_test}, we plot the Bo{\"o}tes-derived SFRD for different choices of lower (left) and upper (right) luminosity limits. We find that our derived SFRD is robust to changes in the lower limit between $0.003\,L_{\star}$ and $0.1\,L_{\star}$ (we adopt $0.03\,L_{\star}$ in this work) and to changes in the upper limit between $10^{27.5}\,\rm{W\,Hz^{-1}}$ and $10^{28.5}\,\rm{W\,Hz^{-1}}$ (we adopt $10^{28}\,\rm{W\,Hz^{-1}}$ in this work).

\bsp	
\label{lastpage}
\end{document}